\documentclass[12pt]{article}

  \newcommand{\onefigure}[2]{\begin{figure}[tbhp]
         \caption{\small #2\label{#1}(#1)}
         \end{figure}}
\renewcommand{\onefigure}[2]{\begin{figure}[tbhp]
    \begin{center}\leavevmode\epsfbox{#1.eps}\end{center}
    \caption{\small #2\label{#1}}
   \end{figure}}

\usepackage{latexsym,amsmath,amssymb,theorem, epsf}

\DeclareFontFamily{OT1}{rsfs10}{}
\DeclareFontShape{OT1}{rsfs10}{m}{n}{ <-> rsfs10 }{}
\DeclareMathAlphabet{\mathscript}{OT1}{rsfs10}{m}{n}

\numberwithin{equation}{section}

\newcommand{\cp}[1]{{\mathbb C}{\mathbb P}^{#1}}

\newcommand{\ns}{\normalsize}

\newcommand{\CC}{{\mathbf{C}}}

\newcommand{\RR}{{\mathbf{R}}}

\def\e{\epsilon}

\def\cC{{\mathcal C}}

\def\cG{{\mathcal G}}

\def\cP{{\mathcal P}}

\def\cS{{\mathcal S}}

\theoremstyle{plain}

\theoremstyle{plain}

\numberwithin{theorem}{section}

\theoremstyle{remark} {\theorembodyfont{\rmfamily}
}

\begin{document}


\begin{titlepage}

\vspace{-5cm}

\title{
   \hfill{\ns UPR-T960, OUTP-99-03P} \\[1em]
   {Knots, Braids and BPS States in M-Theory} \\[1em] }
\author{
   Antonella Grassi $^1$, Zachary Guralnik$^3$
   and  Burt A.~Ovrut $^2$ \\[0.5em]
   {\ns $^1$ Department of Mathematics, University of Pennsylvania} \\[-0.4em]
   {\ns Philadelphia, PA 19104--6395, USA}\\
   {\ns $^2$ Department of Physics, University of Pennsylvania} \\[-0.4em]
   {\ns Philadelphia, PA 19104--6396, USA}\\
   {\ns $^3$ Institut fur Physik, Humboldt Universitat}\\[-0.4em]
{\ns Invalidenstrasse 110, 10115 Berlin, Germany}}

\date{}

\maketitle

\begin{abstract}
In previous work we considered $M$-theory five branes wrapped on
elliptic Calabi-Yau threefold near the smooth part of the
discriminant curve. In this paper, we extend that work to compute
the light states on the worldvolume of five-branes wrapped on
fibers near certain singular loci of the discriminant. We regulate
the singular behavior near these loci by deforming the
discriminant curve and expressing the singularity in terms of
knots and their associated braids. There braids allow us to
compute the appropriate string junction lattice for the
singularity and,hence to determine the spectrum of light BPS
states. We find that these techniques are valid near singular
points with $N=2$ supersymmetry.

\end{abstract}

\thispagestyle{empty}

\end{titlepage}


\section{Introduction}


In this paper, we will consider elliptically fibered Calabi--Yau
threefolds, $X$, over base surfaces, $B$. The discriminant curve
is composed of smooth regions as well as  singular points. For
specificity, in \cite{ggo} the discriminant curves associated with
some Calabi--Yau elliptic fibrations over the blown--up Hirzebruch
base space $B=\hat{\mathbb F}_{3}$ were computed. The fibers of an
elliptic fibration are smooth except over the discriminant locus,
where they degenerate in specific ways classified in part by
Kodaira \cite{kod}. There are also non-Kodaira fibers which may
occur over singularities of the discriminant curve.  In general,
the Kodaira type of the fiber degeneration may be different over
different smooth components of the discriminant locus. In
\cite{ggo}, we discussed the Kodaira classification of
degenerating elliptic fibers and explicitly computed the Kodaira
type of fibers over the discriminant curves associated with the
base $B=\hat{\mathbb F}_{3}$. An example of such a discriminant
curve, indicating the Kodaira type of fiber degeneration, is shown
in Figure $1$.

\onefigure{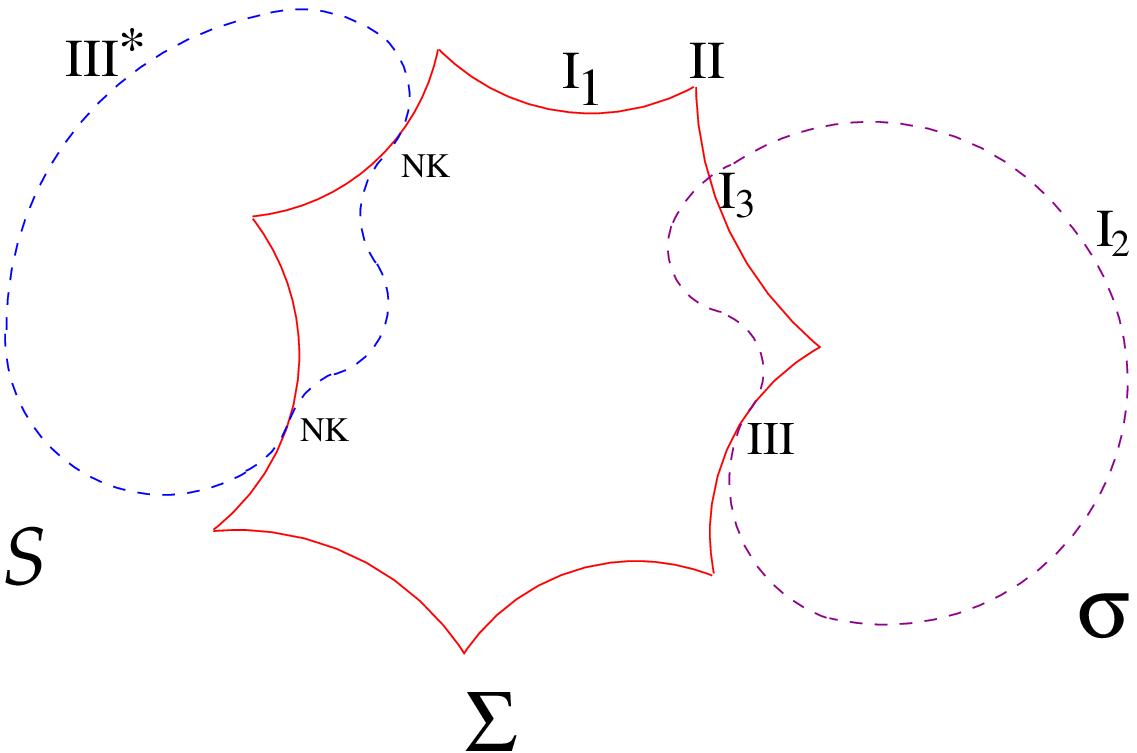}{Schematic illustration of a discriminant
curve.
        The fibers over the smooth parts of the curves $\cS$, $\sigma$ and $\Sigma$
        are of Kodaira type $III^{*}$, $I_{2}$ and $I_{1}$ respectively. The
        fibers over
        the cusp points of $\Sigma$ are of Kodaira type $II$.
        There are non-Kodaira fibers (NK) over the points where the
        $\cS$ component intersects the $\Sigma$ components normally. There are
        $I_3$ fibers where the $\sigma$ component meets the $\Sigma$ component
        normally, and fibers of type $III$ where the $\sigma$ component
        intersects the $\Sigma$ component tangentially.}

It was shown in a series of papers \cite{losw1, nse, don1,fbs, ms,
ppm, si} that, when Ho\v rava--Witten theory \cite{HW1} is
compactified on an elliptically fibered Calabi--Yau threefold, the
requirements that there be three generations of quarks and leptons
on the ``observable brane'' and that the theory be anomaly free
generically necessitates the existence of wrapped M five--branes
in the five--dimensional ``bulk space''. These five--branes have
two space--like dimensions wrapped on a holomorphic curve in the
Calabi--Yau threefold. In \cite{ggo} and in this paper, we are be
interested in the case when this holomorphic curve is a pure
fiber, $\cC_{2}$. The five--brane worldvolume manifold is then
$M_{4} \times \cC_{2}$. The four dimensional theory arising from
the wrapped five--brane has $N=1$ supersymmetry.  However the
amount of supersymmetry may be enhanced at low energies, depending
on the five--brane location in the base. For $\cC_{2}$ located at
a generic point in the base, it can be shown that, at low energy,
the $M_{4}$ worldvolume theory always contains a single $U(1)$
$N=4$ Abelian vector supermultiplet. However, if the fiber
approaches any point on a smooth component of the discriminant
curve, the $M_{4}$ worldvolume supersymmetry is enhanced to $N=2$.
In addition to the $N=2$ decomposition of the $N=4$
supermultiplet, new light BPS hypermultiplets carrying $U(1)$
charges appear. The states may carry mutually non-local dyonic
charges,  which can not simultaneously be made purely electric by
an $SL(2,Z)$ transformation. In that case, these theories are
exotic and do not have a local Lagrangian description.  When the
five--brane wraps the degenerate fiber over the smooth component
of the discriminant curve,  the theory flows to an $N=2$
interacting fixed point in the infrared,  of a type originally
described (in a different context) in \cite{argyresdouglas,
argwitten, minahan}. In \cite{ggo}, we considered only the smooth
regions of the discriminant locus and using the theory of string
junction lattices \cite{bartonone, bartontwo} we presented strong
constraints on the spectrum and the local and global charges of
the additional light BPS hypermultiplets on $M_{4}$.

The discussion in \cite{ggo} was restricted to smooth regions of
the discriminant locus for two reasons. First, the supersymmetry
at such points is $N=2$,  and there are powerful constraints on
the possible BPS multiplets \cite{bartontwo}. Furthermore,  the
degenerate fibers over a smooth component of the discriminant
curve fall entirely under Kodaira's classification, and there is
no ambiguity in the computation of the BPS states. However,
neither of these is always a property of singularities of the
discriminant locus,  as we will see below.

The supersymmetry on a five--brane wrapped near such points is not
always enhanced beyond $N=1$ at low energies.
In these cases,  a theory of a five brane wrapping a
degenerate fiber over the singular point is an $N=1$ fixed point
theory of the type discovered in \cite{aks}. In \cite{aks} these
theories appeared in the context of three-brane probes of F-theory
compactifications on elliptic Calabi-Yau threefolds. In
\cite{aks},  singularities of the discriminant locus giving rise
to such singularities were identified,  and certain anomalous
dimensions were obtained. Like its $N=2$ counterpart, the low
energy theory on a five-brane wrapped near a singularity of the
discriminant curve is expected to have matter with mutually
non-local charges, however much less is known about the spectrum.
In this paper we begin a study of the spectrum near singularities
of the discriminant curve.

In \cite{ggo} we deformed the Weierstrass model so that the new
discriminant curve was a local, disjoint union of $I_1$ curves. We
showed  that this method can be successfully applied to points on
the smooth parts of the discriminant curve. In this paper, we
extend these results to elliptic fibrations in a neighborhood of a
singular locus of the fibration.

One could attempt to deform the Weierstrass model around a
singular point $P_0$ of the discriminant in such a way that the
new discriminant curve is a locally smooth curve with $I_1$
fibers. Deformations of this type could, in principle, be applied
to some singular discriminant loci, albeit in a manner that
differs significantly from the smooth case. Even in these cases,
however, technical difficulties arise. We are motivated,
therefore, to present some other procedure to ``regulate'' the
singular loci. In other cases it is simply not possible to deform
the isolated singularity of the discriminant.

Our approach will be the following: first, we will choose certain
deformations of the Weierstrass model, so that the discriminant
locus will still have the same multiplicity  at $P_0$, and the
 nearby singular fibers are of Kodaira type $I_1$.  We then ``regulate" the singular point $P_0$
of the discriminant by surrounding it with an infinitesimal real
sphere, $S^{3}$. A similar approach appeared in
\cite{argyresdouglas},  in the context of $SU(3)$ ${N} =2$
Yang-Mills theory. The intersection of the discriminant curve with
this sphere produces either a knot or a link in $S^3$ which is
characteristic of the type of the singularity at $P_0$. All the
points on this knot or link are within a smooth component of the
discriminant curve of Kodaira type $I_1$. Any knot or link can be
represented, but not uniquely, by a braid and an associated
element of the braid group.

Next, we  choose a braid and  we seek the equivalence classes of
string junctions on $S^3$,  that is, the classes of junctions
related by deformations and by Hanany-Witten transitions. Now,
using the braid group and Hanany--Witten transformations, one can
show that all possible $S^{3}$ string junction configurations are
equivalent to a string junction in a real surface ${\cal{P}}$
transverse to the braid.
 This is possible because  the allowed deformations
do not change the singularity type of the general surface through
$P_0$.

Since intersection number can be defined for membranes in the
elliptic fibration over this surface, it follows that one can
define the associated string junction lattice exactly as in
\cite{ggo, bartonone}.  However, there remains an ambiguity that
must be resolved. This arises from the fact that the braids
associated with a given knot or link are not unique. We resolve
these ambiguities by showing, using the braid group and
Hanany--Witten transformations, that the string junction lattices
associated with any two braid representations arising in this way
are equivalent.
 In fact, we will show that
the choice of an elliptic surface   in the neighborhood of the
singularity is related to the choice of a braid representative.
 The ``general elliptic" surface corresponds to the ``minimal braid"
(defined in A.2);  when our method can be applied, the junction
lattice of the threefold is that of the general elliptic surface
through $P_0$.

 Our method cannot always be  successfully applied.
   In Section \ref{susy1} we
   give one such example:   the
 supersymmetry here is broken to ${N} =1$ at the point $P_0$. In Section \ref{susy2} we
  discuss the relationship with
   supersymmetry breaking.

  \medskip

  The outline of the paper is as follows:
  in Section \ref{methods} we state the procedure;
  in Section \ref{notation} we recall basics facts about elliptic fibrations,
singular fibers and discriminants.
  In Section \ref{cusps} we consider an example where the discriminant has a singularity at
 a point $P_0$ and the nearby singular fibers are of Kodaira type
 $I_1$. In this case we do not need to deform the Weierstrass
 equation; instead  we explain in details how two different
 braids give the same equivalence class of string junction
 lattices. In the Appendix, we explicitly describe  the knot
 associated with this discriminant singularity and the two different braids
 in consideration.

In Section \ref{def}, we work out three examples where the
discriminant has to be  deformed. These examples illustrate the
various rules stated in Section \ref{methods}. It is worth noting
that we can also recover the results
 of the previous paper \cite{ggo}, albeit by a less direct
 procedure. To illustrate this, in the first example of
 \ref{def} we apply our procedure  in a neighborhood of a point in the smooth part
of the discriminant with Kodaira type $I_2$.

\medskip

Remark:  The order of vanishing of the discriminant of the
threefold at $P_0$ coincides with the multiplicity of the
discriminant of the general elliptic surface through $P_0$. The
latter is the number of end points in the surface string junction
lattice, which equals the number of strands in the minimal braid
(see A.1 and A.2). This made us suspect that the multiplicity of
an isolated curve singularity would always coincide with the
number of strands of the minimal braid (also called the ``index"
of the knot, or link). It turns out that this statement is indeed
always true; its proof is a hard and beautiful result. The first
proof was given in \cite{schubert}. A shorter, elegant proof is
given by \cite{williams}, using techniques from dynamical systems
(see also \cite{libgober} in the same volume, for an algebraic
version). For an outline of the proof in the case of`` torus
knots'' (see the Appendix and, for example \cite{murasugi}).

\section{Outline of our procedure}\label{methods}

First, we will choose a deformation of the Weierstrass model
around $P_0$, which fixes the order of vanishing of the
discriminant at the origin, so that the other nearby singular
fibers become of Kodaira type $I_1$. This imply that the general
elliptic surface through the $P_0$ of the deformed model has the
same type of singularity over $P_0$ as in the non-deformed model.

 We
then ``regulate" the singular point $P_0$ of the discriminant by
surrounding it with an infinitesimal real sphere, $S^{3}$. The
intersection of the discriminant curve with this sphere produces
either a knot or a link in $S^3$ which is characteristic of the
type of the singularity at $P_0$ (see the Appendix).
 All the points
on this knot or link are within a smooth component of the
discriminant curve of Kodaira type $I_1$.

Next, we  choose a braid representative of this knot (link) (see
Appendix A) and we seek the equivalence classes of string
junctions on $S^3$, that is, the classes of junctions related by
deformations and by Hanany-Witten transformations. Note that there
is nothing which forces junctions to live on this $S^3$. Actually
they live in a four ball $B^4$ surrounding the singular point.
However,  we will assume that the equivalence classes of string
junctions on $B^4$ are the same as the equivalence classes on $B^4
- P_0$,  where $P_0$ is the singular point. The endpoints of the
junctions can then be slid along the discriminant curve, so that
they lie entirely on the $S^3$ boundary of $B^4$. Now, using the
braid group and Hanany--Witten transformations, one can show that
all possible $S^{3}$ string junction configurations are equivalent
to a string junction in a real surface ${\cal{P}}$ transverse to
the braid.
 This is possible because  the allowed deformations
do not change the singularity type of the general surface through
$P_0$.

Since intersection numbers can be defined for membranes in the
elliptic fibration over this surface, it follows that one can
define the associated string junction lattice exactly as in
\cite{ggo, bartonone}.  However, there remains an ambiguity that
must be resolved. This arises from the fact that the braids
associated with a given knot or link are not unique. We resolve
these ambiguities by showing, using the braid group and
Hanany--Witten transformations, that the string junction lattices
associated with any two braid representations arising in this way
are equivalent. Furthermore  we show that the string junction
lattice in the threefold is the string junction lattice of ``the
minimal braid" associated to the knot(link) (see  A.2 for the
definition). The braid is obtained by ``cutting" the discriminant
with a (complex) curve $C_R$, which depends on the radius $R$ of
the $S^3$; in the limit as $R \to 0$, $C_R$ becomes a general
curve $C$ through $P_0$ (see A.3).

  In our examples, the string junction lattice of the elliptic surface over $C$
 coincides with the string junction lattice of the
 threefold.  We speculate that this will not happen in
 more general cases, where the supersymmetry at
 $P_0$ is broken at ${ N}=1$.

\smallskip

Note that with other deformations of the equation  which do not
satisfy our requirements, it is no longer clear how to understand
a ``cut'' of this braid as a deformation of the original Kodaira
type.
 This is because the requirement implies that
  the {\it index} (i.e. the minimal number of strands
  of an associated braid) of the knot (link) will stay the same
  before and after the ``allowed deformations".
This can be explained as follow: On an elliptic surface,  the
string junction lattice is obtained by locally deforming the
Weierstrass  equation so as to split Kodaira fibers into fibers of
$I_1$ type.  The number of $I_1$ fibers into which a Kodaira fiber
splits is determined by the order of vanishing of the
discriminant. There are deformations  which introduce more $I_1$
fibers entering from infinity, but none which increase the number
locally. {(In cases in which the Kodaira fiber corresponds to a
conformal field theory, the deformations which introduce new $I_1$
fibers at infinity correspond to irrelevant deformations.)}  There
is thus  no smooth deformation which increases the number of
states.

On an elliptically fibered Calabi-Yau,  the string junction
lattice may again be defined by splitting the fibers in the
neighborhood of a singularity of the discriminant curve into $I_1$
type. The dimension of the string junction lattice is then
obtained from the braid index of the resulting knot or link.  As
for a K3,  we shall assume that there is no smooth deformation
which increases the number of states by raising the dimension of
the string junction lattice, (although in this case there may be
deformations which lower the dimension.)

\section{Basic facts about elliptic fibrations}\label{notation}

A simple representation of an elliptic curve is given in the
projective space $\cp{2}$ by the Weierstrass equation
\begin{equation}
zy^2=4x^3-g_2xz^2-g_3z^3 \label{eq:31}
\end{equation}
where $(x,y,z)$ are the homogeneous coordinates of $\cp{2}$ and
$g_2$, $g_3$ are constants. The origin of the elliptic curve is
located at $(x,y,z)=(0,1,0)$. The torus described by~\eqref{eq:31}
can become degenerate if one of its cycles shrinks to zero. Such
singular behavior is characterized by the vanishing of the
discriminant
\begin{equation}
\Delta= g_{2}^{3}-27g_{3}^{2} \label{eq:31A}
\end{equation}

Equation~\eqref{eq:31} can also represent an elliptically fibered
surface (or threefold), $W$, if the coefficients $g_2$ and $g_3$
in the Weierstrass equation are functions over a base curve  (or
surface) $B$ (see for example \cite{ggo}, for more details).

The resolved singular fibers of a Weierstrass model representing
an elliptic curve are determined by the order of vanishing of
$g_2, g_3$ and $\Delta$; these fibers were classified by Kodaira:

\vskip 0.2in

\begin{tabular}{|l|c|c|c|}
\hline
Kodaira type & A-D-E & monodromy & N,L,K \\
\hline $I_n$ & $A_{n-1}$ & $\begin{pmatrix} 1 &  n \\ 0 &  1
\end{pmatrix} $ &
$N=n, L=0, K=0$ \\
\hline
$II$ & & $\begin{pmatrix}1 & 1\\ -1 & 0 \end{pmatrix}$ & ${N=2, L>0, K=1}$ \\
\hline $III$ & $A_1$ & $\begin{pmatrix}0 & 1\\ -1 &
0\end{pmatrix}$ & $N=3, L=1,
K>1$\\
\hline $IV$ & $A_2$ & $\begin{pmatrix}0 & 1\\ -1 & -1
\end{pmatrix}$ & $N=4, L>1,
K=2$\\
\hline $I_0^*$ & $D_4$ & $\begin{pmatrix} -1 & 0\\ 0 & -1
\end{pmatrix}$ & $N=6,
L>1, K>2$\\
\hline $I_n^*$ & $D_{n+4}$ & $\begin{pmatrix}-1 & -n \\ 0 &
-1\end{pmatrix}$ & $N
= 6+n, L=2, K=3$\\
\hline $IV^*$ & $E_6$ & $\begin{pmatrix} -1 & -1 \\ 1 & 0
\end{pmatrix}$ & $N = 8,
L>2, K=4$\\
\hline $III^*$ & $E_7$ & $\begin{pmatrix} -0 & -1 \\ 1 &
0\end{pmatrix}$ & $N = 9,
L=3, K>4$\\
\hline $II^*$ & $E_8$ & $\begin{pmatrix} 0 & -1 \\ 1 & 1
\end{pmatrix}$ & $N = 10,
L>3, K=4$\\
\hline
\end{tabular}

\vspace{30pt}

{\small \noindent Table 1: The integers $N$, $L$ and $K$
characterize the behavior of $\Delta$, $g_{2}$ and $g_{3}$ near
the discriminant locus $u=0$; $\Delta = u^Na , g_{2} = u^Lb$ , and
$g_{3} =u^Kc$ .}

\section{Regularizing without deforming the equation: a general cusp singularity of the discriminant}\label{cusps}

Isolated cusp points are a generic feature of discriminant curves.
For example, in the discriminant locus shown in Figure $1$, there
are $137$ cusps in the $\Sigma$  component of the curve. Note that
any deformation of the Weierstrass equation corresponds to a
change of coordinates in the $(s,t)$ plane. Thus it is never
possible to remove the singularity of the discriminant by any
deformation.

Consider one such cusp and choose local coordinates $s,t$ which
vanish at that point. Then, in the neighborhood of the cusp, the
sections $g_{2}$ and $g_{3}$, which define the elliptic fibration,
can be taken to be
\begin{equation}
g_{2}=s,  \qquad g_{3}=t \label{eq:1A}
\end{equation}
It follows that the Weierstrass representation of the fiber and
the discriminant are given by

\begin{equation}
y^{2}=4x^{3}-sx-t \label{eq:2}
\end{equation}
and
\begin{equation}
\Delta= s^{3}-27t^{2} \label{eq:1}
\end{equation}
respectively.

Note that the Weierstrass model $y^2 = x^3 - t$ over the surface
$s=0$ appears to be that of a type II Kodaira fiber at $t=0$, but
the Weierstrass model $y^2 = x^3 - sx$ over the surface $t=0$
appears to be that of a type $III$ Kodaira fiber at $s=0$. To
discuss the monodromy around a point in the discriminant locus of
an elliptically fibered threefold, we could restrict to the
fibration over a curve that intersects the discriminant locus at
the cusp.  The surface over generic curves, however, are smooth
everywhere: note that the path $\cP$ defined by $s=0$, which has
coordinate $t$, is generic. For specificity, we take the
intersecting surface to be the elliptic fibration over $\cP$,
which we denote by ${\cal{T}}$. Restricted to this surface, the
cusp appears as a point in the one--dimensional complex base
$\cP$. Furthermore,  restricted to ${\cal{T}}$, the discriminant
and the Weierstrass representation are given by
\begin{equation}
\Delta=-27t^{2},  \qquad y^{2}=4x^{3}-t \label{eq:3}
\end{equation}
Both of these quantities are recognized as corresponding to a
Kodaira type $II$ degeneration of the fiber at the cusp.  Since
the problem has now been reduced to an elliptic surface over a
one--dimensional base, we can apply standard Kodaira theory to
analyze the monodromy.

The $SL(2, {\mathbb Z})$ monodromy transformation for a Kodaira
type $II$ fiber is given by
\begin{equation}
{\cal{M}}_{II}=\begin{pmatrix}\
                  1 & 1 \\
                  -1 & 0
                  \end{pmatrix}.
\label{eq:4}
\end{equation}
This transformation acts on the elements of $H_{1}(\cC_{2},
{\mathbb Z})$, where $\cC_{2}$ is the elliptic fiber. Note that
${\cal{M}}_{II}$ has no real eigenvector and, therefore, no
obvious associated vanishing cycle. The usual way to proceed is to
use the fact that ${\cal{M}}_{II}$ can be decomposed as
\begin{equation}
{\cal{M}}_{II}= \begin{pmatrix}
                  1 & 1 \\
                  0 & 1
                  \end{pmatrix}
\cdot           \begin{pmatrix}
                  2 & 1 \\
                  -1 & 0
                  \end{pmatrix}
=A \cdot B \label{eq:5}
\end{equation}
where both matrices $A,B$ are monodromies of Kodaira type $I_{1}$.

However, we have to justify the choice of the ``generic" elliptic
surface through $P_0$. It is in fact easy to see that non-generic
surfaces will give different string junctions.

In order to resolve this ambiguity, following  \cite{argyresdouglas}, we
will consider, not the cusp itself, but the intersection of the
discriminant curve
\begin{equation}
s^{3}-27t^{2}=0
 \label{eq:6}
\end{equation}
with a l sphere $S^{3}$ of infinitesimal radius $R$
centered at the cusp.
We will see in  A.3 that the intersection is the ``trefoil'' knot
shown in Figure $2$. All points in the knot lie in the smooth part
of the discriminant curve, over which the elliptic fiber is of
Kodaira type $I_1$, as shown in Figures $1$ and $2$.

\onefigure{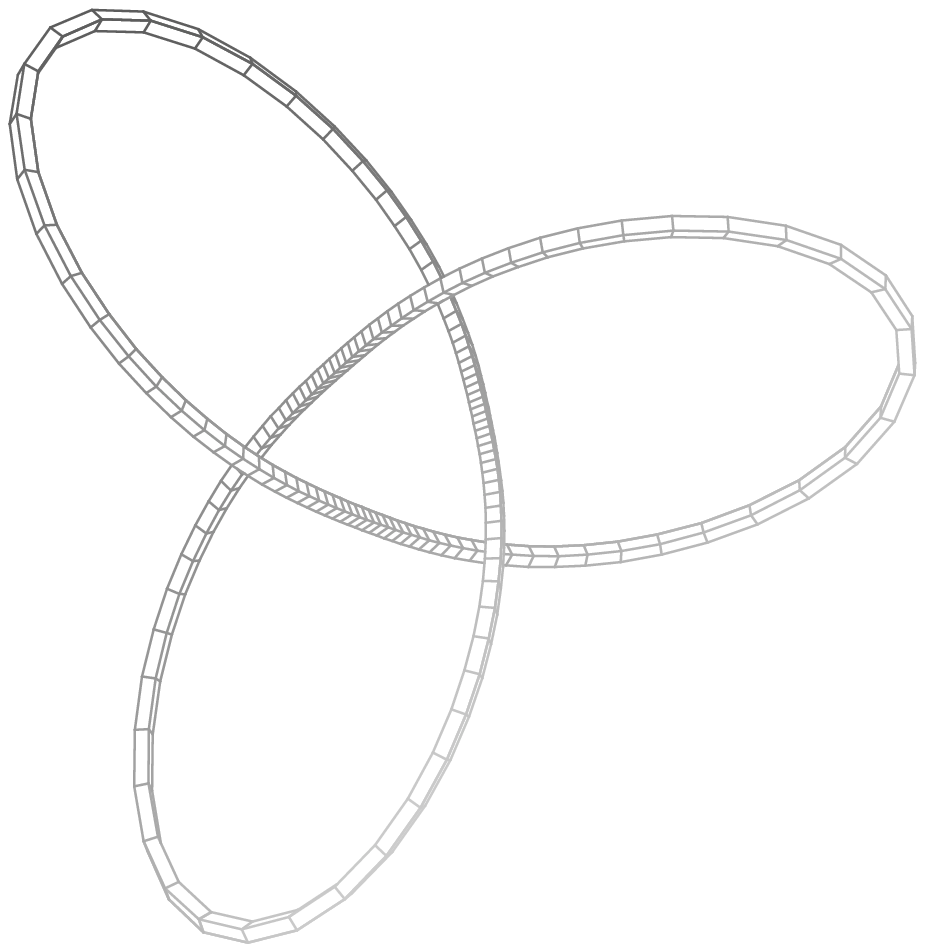}{The trefoil knot associated with a cusp
singularity. There is a Kodaira type $I_1$ fiber over every point
in the knot.}

To proceed, it is useful to represent the trefoil knot as a braid.
This representation is not unique, as we will see in the Appendix.
Two relevant braid representatives of the trefoil knot are shown
in Figure $3$; we derive them explicitly in  A.3.

\onefigure{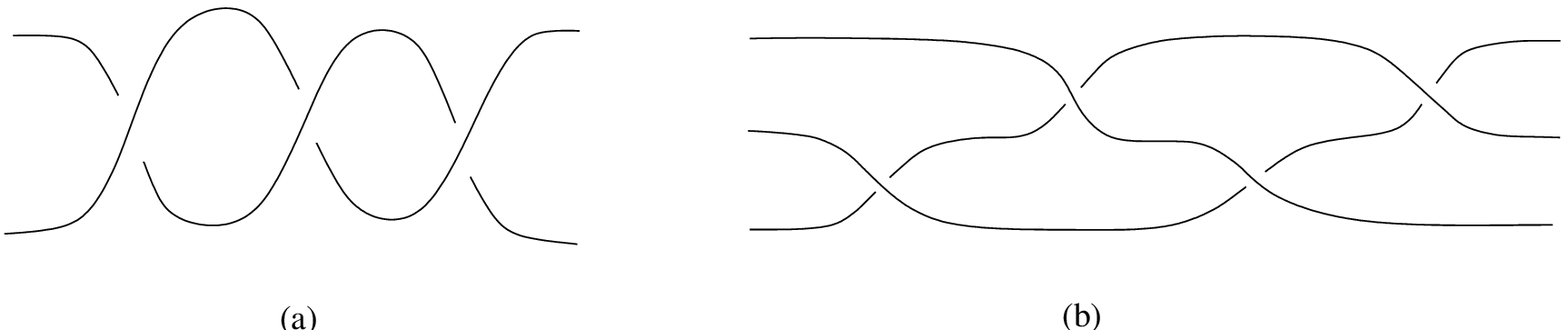}{Two equivalent braid representations of the
trefoil knot (see also figure \ref{cuspinbraids}).
 The
opposing endpoints are identified.}

Note that any braid representative is a curve in ${\RR}^{3}$.
First, consider the two strand braid shown in Figure $3(a)$. This
is obtained by cutting the knot with the line $s=c_R$, where $c_R$
is a constant depending on $R$ and $c_R \to 0$, as $R\to 0$ (see
A.3). In this limit, this is one of the generic lines through the
origin, intersecting the cusp with multiplicity $2$.
 Fix some arbitrary point $P$ in ${\mathbb R}^{3}$
(corresponding to the location of the wrapped M5-brane) which is
not on the braid, and specify an ${\mathbb R}^{2}$ plane, $\cP$,
which contains $P$ and intersects the braid. This is shown in
Figure $4(a)$. Note that the braid intersects $\cP$ in two points.
The elliptic fibration over $\cP$, which we denote by
${\cal{T}}_{R}$, is our regularization of the surface ${\cal{T}}$.
${\cal{T}}_{R}$ has two separated discriminant points, each of
Kodaira type $I_{1}$. In an appropriate basis,  the monodromies
around these two points in the plane $\cP$ are $A$ and $B$ as
defined in equation (2.6),  as illustrated in figure $4(a)$. Note
that in the planes over neighboring sections of the braid,  the
monodromies are different.  The relation between monodromies in
different sections of a braid is shown in Figure $5$. Next, extend
a single string from $P$ to an arbitrary point on the braid which
is not, in general, in $\cP$. This point can always be moved back
into the plane using a generalized Hanany--Witten mechanism. The
result is a two--legged string junction that can be made to lie
entirely in the $\cP$ plane. This is shown in Figure $4$.

\onefigure{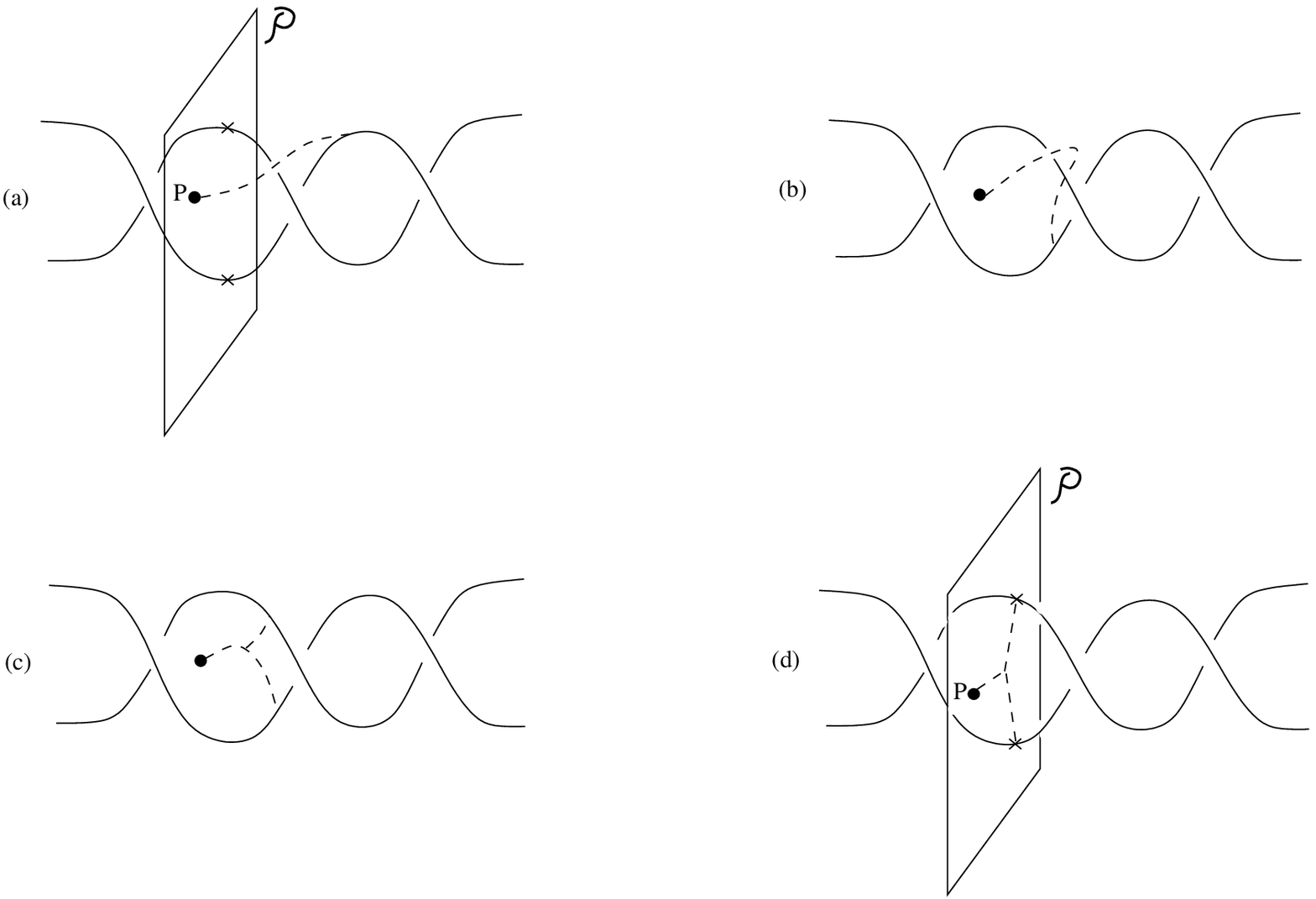}{Mapping string junctions into a plane. By
sliding the endpoints of the junction and doing Hanany--Witten
transformations, as between Figures (b) and (c), any string
junction can be brought into the plane, as in Figure (d).}

\onefigure{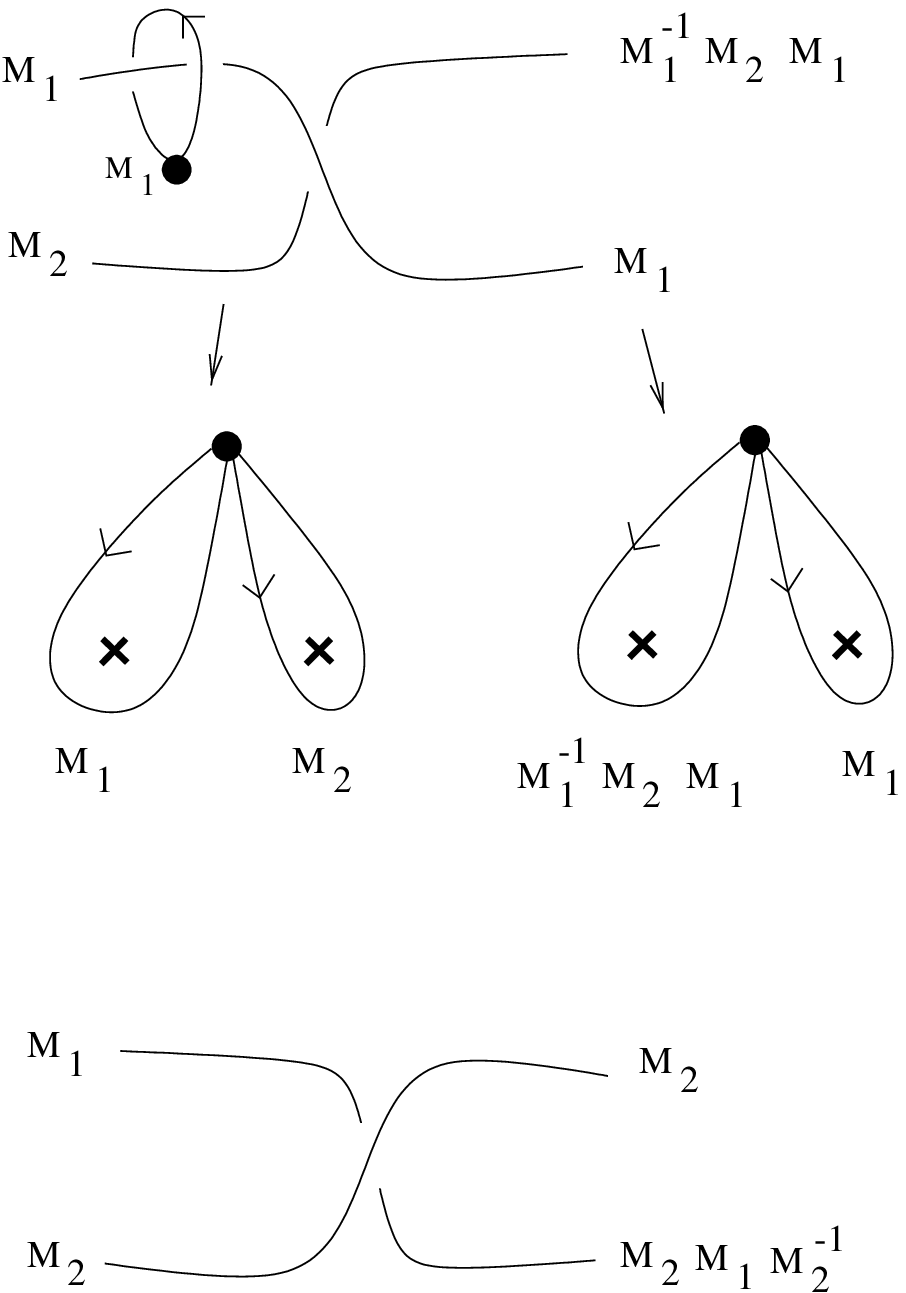}{Illustration of the relation between
$SL(2,Z)$ monodromies in different sections of a braid.}

It is not hard to show that any string junction starting at $P$
can always be represented by a two--legged string junction in the
plane $\cP$. Note that the string junction lattice we obtain in
this manner corresponds to a Kodaira type $II$ fiber degeneracy.
Having put a junction in the plane,  one can then slide  endpoints
of the junction lattice all the way around the braid and then do
Hanany-Witten transitions to bring the junction back into the
plane.  This could, in principle, generate an additional
equivalence relation,  equating apparently different junctions in
the plane $\cP$.  It is easy to see that, in this case, there is
no such equivalence relation. This is because the total $(p,q)$
charge of a junction (or the boundary cycle of the membrane which
results from lifting to M-theory) is not changed by this process.
Since there are only two $I_1$ loci in the plane, the $p$ and $q$
charges completely determine the equivalence class of a junction
in the plane (defined up to deformations and Hanany-Witten
transitions within the plane).  Thus,  in this instance, the
equivalence classes of junctions are the same as the equivalence
classes of junctions restricted to the plane $\cP$.

Before discussing these states, however, we would like to point
out that one could have considered the surface over the line $t=0$.
This line intersects the discriminant curve with multiplicity $3$.
The path $\cP$, with coordinate $s$, is  here defined by $t=0$. To
begin, denote the elliptic fibration over $\cP$ by ${\cal{T}}$ and
note that the cusp appears as a point in the one--dimensional
complex base. Furthermore, the discriminant and Weierstrass
representation are given by
\begin{equation}
\Delta=s^{3}, \qquad y^{2}=4x^{3}-sx \label{eq:9}
\end{equation}
which is the Weierstrass representation of a type $III$ Kodaira
fiber in an elliptic surface. Since the problem has now been
reduced to an elliptic two-fold, we can apply standard Kodaira
theory to analyze the monodromy.

The $SL(2,{\mathbb Z})$ monodromy transformation for a Kodaira
type $III$ fiber is given by
\begin{equation}
{\cal{M}}_{III}= \begin{pmatrix}\
                    0 & 1 \\
                   -1 & 0
                   \end{pmatrix}
                  = A \cdot\ A \cdot B
\label{eq:9A}
\end{equation}
where matrices $A, B$, defined in~\eqref{eq:5}, both correspond to
monodromies of Kodaira type $I_{1}$. Since ${\cal{M}}_{III}$ has
no real eigenvector, one can attempt to proceed by deforming the
discriminant curve from a single point of Kodaira type $III$ to
three separate points, each of Kodaira type $I_{1}$. This
deformation corresponds to the Weierstrass model of equation
 $$  y^{2}=4x^{3}-sx + t.$$
Here $t$ is the deformation parameter; the three $I_1$ fibers are
found by considering the plane $t=c_R$, where $c_R$ is a constant
(which depends on the radius $R$). These three singular fibers are
exactly the ones obtained by a ``vertical cut'' (with the line
$t=c_R$) in the 3-braided representation of the trefoil knot, as
shown in Figure 3(b) (compare also with A.3).

Choose some point $P$  (corresponding to the location of the
wrapped M 5-brane) which is not on the braid and specify a plane
$\cP$ which passes through $P$ and intersects the braid. This is
depicted in Figure $6$. Note that the braid intersects $\cP$ in
three points. The elliptic fibration over $\cP$, which we denote
by ${\cal{T}}_{R}$, is our regularization of the surface
${\cal{T}}$. ${\cal{T}}_{R}$ has three separated discriminant
points, each with Kodaira type $I_{1}$. The monodromies of each of
these points can be computed and in an appropriate basis are the
matrices $A$, $A$, and $B$ respectively in the plane $\cP$
indicated in Figure 6.

By sliding string endpoints along the braid and using the
Hanany--Witten mechanism, it can be shown that any string junction
starting on $P$ and ending at arbitrary points on this braid can
always be deformed to a three-legged string junction lying
entirely within the plane $\cP$.

\onefigure{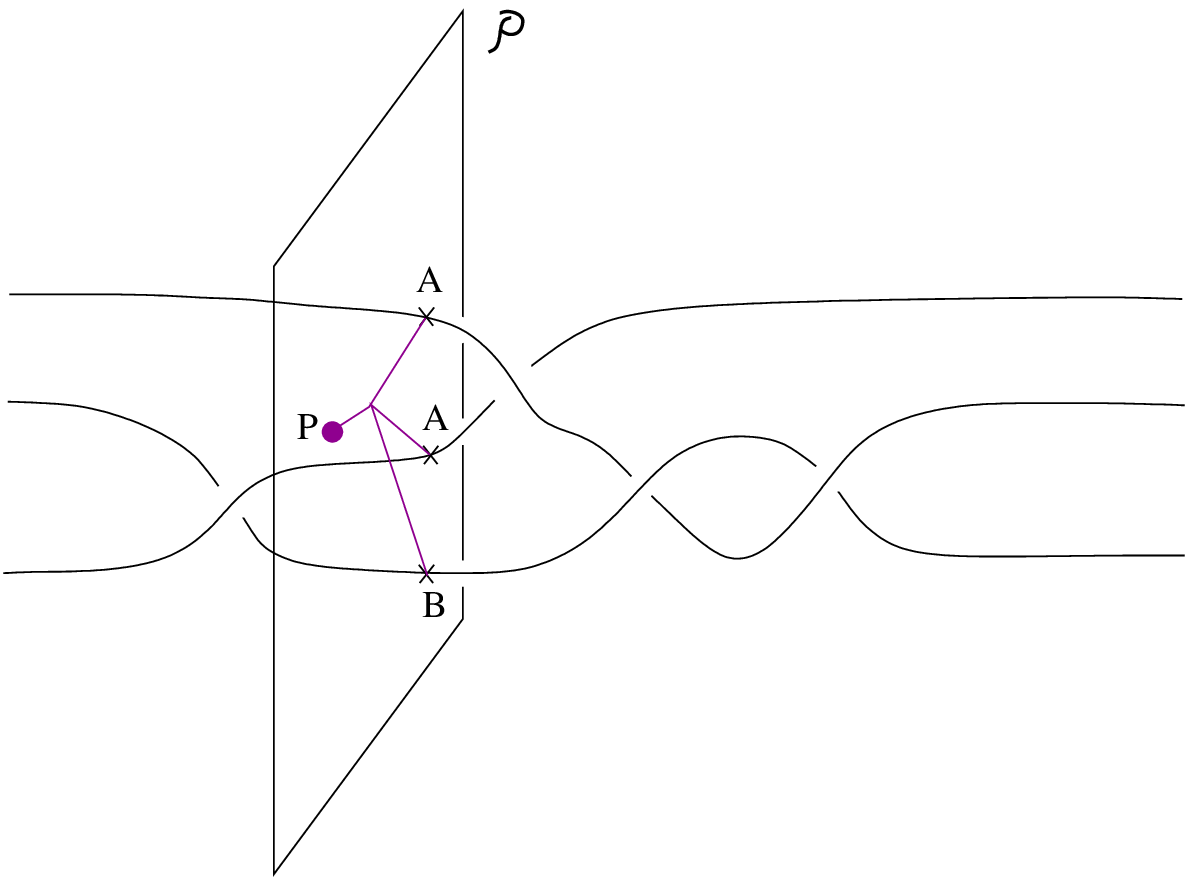}{String junctions can always be brought
into the plane intersecting the three--segment braid
representation of the trefoil knot.}

Note that the string junction lattice we obtain from this
three--strand  braid representation appears to be one associated
with a Kodaira type $III$ fiber degeneracy and, hence, in
contradiction with the above result.  However, despite
appearances, the string junction lattice associated with the
three--strand braid is equivalent to the two--strand braid string
junction lattice. Generically, the reason for this is the
following. Unlike the type $III$ string junction lattices at a
smooth point in the discriminant curve \cite{ggo, bartonone}, we
can define a set of transformations where either one, two or all
three legs of the junction can be translated along the braid until
they return to the plane $\cP$. As a rule, one will obtain a
different string junction in the plane $\cP$. However, by
construction, this new junction must be equivalent to the original
one. That is, not all Kodaira type $III$ string junctions near a
cusp are independent. We now proceed to show that, in fact, only a
Kodaira type $II$ subset of string junctions  are independent.

The braid under consideration is divided into sections, each with
three parallel lines. An element of the braid group acts between
each section by intertwining the three lines.  As one moves the
junction from one section to the next,  one can do a Hanany-Witten
transition to keep it in a canonical form within a plane
intersecting the braid section,  in which the junction lies in the
upper half-plane. In this way, the braid group acts on string
junctions restricted to the plane.  As one goes all the way around
the braid, string junctions $J$ in the plane $\cP$ are acted on by
the (three-strand) braid group element of the knot $\cG$,  giving
another string junction $J^{\prime}$ which is equivalent to $J$
\begin{equation}
\cG J = J^{\prime} \sim J
\end{equation}
Such an equivalence would not exist if we were considering the
junction lattice of a type $III$ fiber in an elliptic two-fold
instead of a three-fold. The string junction lattice is thus the
quotient of the string junction lattice restricted to the plane
$\cP$ by the action of the braid group. To see explicitly how the
quotient acts, consider an oriented string, denoted by
$\vec{\alpha}$, connecting the two discriminant points in the
$\cP$ plane in Figure $6$ with identical monodromy matrix $A$.
Translating this string along the braid, we find that it returns
to the plane $\cP$ with its orientation reversed, as illustrated
in Figure $7$. We have indicated in the figure the canonical form
of the junction which arises as one moves around the braid. Within
each section of the braid,  the junction may be written as a
vector $\vec{\alpha} = n_1 {\vec v}_1 + n_2 {\vec v}_2 + n_3 {\vec
v}_3$,  where ${\vec v}_i$ is a single string with an endpoint on
the i'th segment of the braid, labelled from top to bottom.  As
one moves around the braid, starting from the point indicated,
$\vec{\alpha}$ is transformed as follows
\begin{equation}
{\vec{\alpha}} = {\vec v}_1 - {\vec v}_3 \rightarrow {\vec v}_1 +
{\vec v}_3 - {\vec v}_2 \rightarrow {\vec v}_3 - {\vec v}_1
\rightarrow {\vec v}_2 + {\vec v}_3 - {\vec v}_1 \rightarrow {\vec
v}_3 - {\vec v}_1 = -{\vec{\alpha}}
\end{equation}
This sequence is illustrated in Figure $7$. It follows that
$\vec{\alpha}$ is equivalent to zero. Therefore,as illustrated in
Figure 8, given any junction we can, by Hanany-Witten
transformations, write it as a two strand junction in the plane
$\cP$ with an endpoint on the strand with monodromy $B$ and an
endpoint on one of the strands with monodromy $A$. There can be no
further equivalence relation because the dimension of the lattice
must be at least two to get all possible $(p,q)$ boundary cycles.
Thus one has the same two dimensional string junction lattice that
one obtains for a type $II$ Kodaira fiber.

Another way of stating the above result is the following. Given
the Weierstrass model $y^2 = 4x^3 -sx -t$,  one can find the
lattice of string junctions restricted to a curve of constant
non-zero $t$. One then quotients by the monodromy which acts on
this lattice as one moves this curve in a circle around $t=0$
\footnote{Note that this monodromy in some ways resembles that
which gives non-simply laced algebras in certain F-theory
compactifications \cite{bershadskyetal, aspinwallkatzmorrison} by
acting on the Dynkin diagram.  However these monodromies are not
equivalent.  In instances in which there are non-simply laced
algebras,  the roots which are projected out by the Dynkin diagram
monodromy have corresponding string junctions which are not
projected out by the monodromy which acts on the junction lattice.
However these junctions give rise to hypermultiplets rather than
vector multiplets.}. This gives the same result as restricting to
curve of constant non-zero $s$, and quotienting by the (in this
case trivial) monodromy which acts when one moves the curve in a
circle around $s=0$.

In this Section, we have only proven the equivalence of the two
and three-braid string junction lattices.  However,  using these
techniques it is not hard to demonstrate this equivalence  for all
braid representations of the trefoil knot. It follows that the
correct structure of the string junction lattice at a cusp,
Kodaira type $II$, is most easily obtained from the two-braid
representation. As discussed above, this corresponds to choosing
the curve intersecting the cusp to be generic.  In general,  it
turns out that the junction lattice in the neighborhood of a
singularity of the discriminant curve is determined by the
$SL(2,Z)$ monodromy around the singular point within a generic
slice. In the present example,  a generic slice containing the
cusp is given locally by $s + \lambda t = 0$ for finite $\lambda$.
The restriction of the Weierstrass equation to this slice is $y^2
= 4x^3 + \lambda t x - t$,  with the discriminant $\Delta = t^2(-
\lambda^3 t - 27)$. The monodromy around $t=0$ is that of a type
$II$ Kodaira fiber. This completely resolves the discrepancy
discussed above.  That is, the junction lattice is that of a type
$II$ Kodaira fiber on an elliptic surface. In fact,  in this case,
the fiber at the  cusp happens to be a type $II$ Kodaira fiber.
Later, we will encounter examples in which the fiber type does not
match the junction lattice.

\onefigure{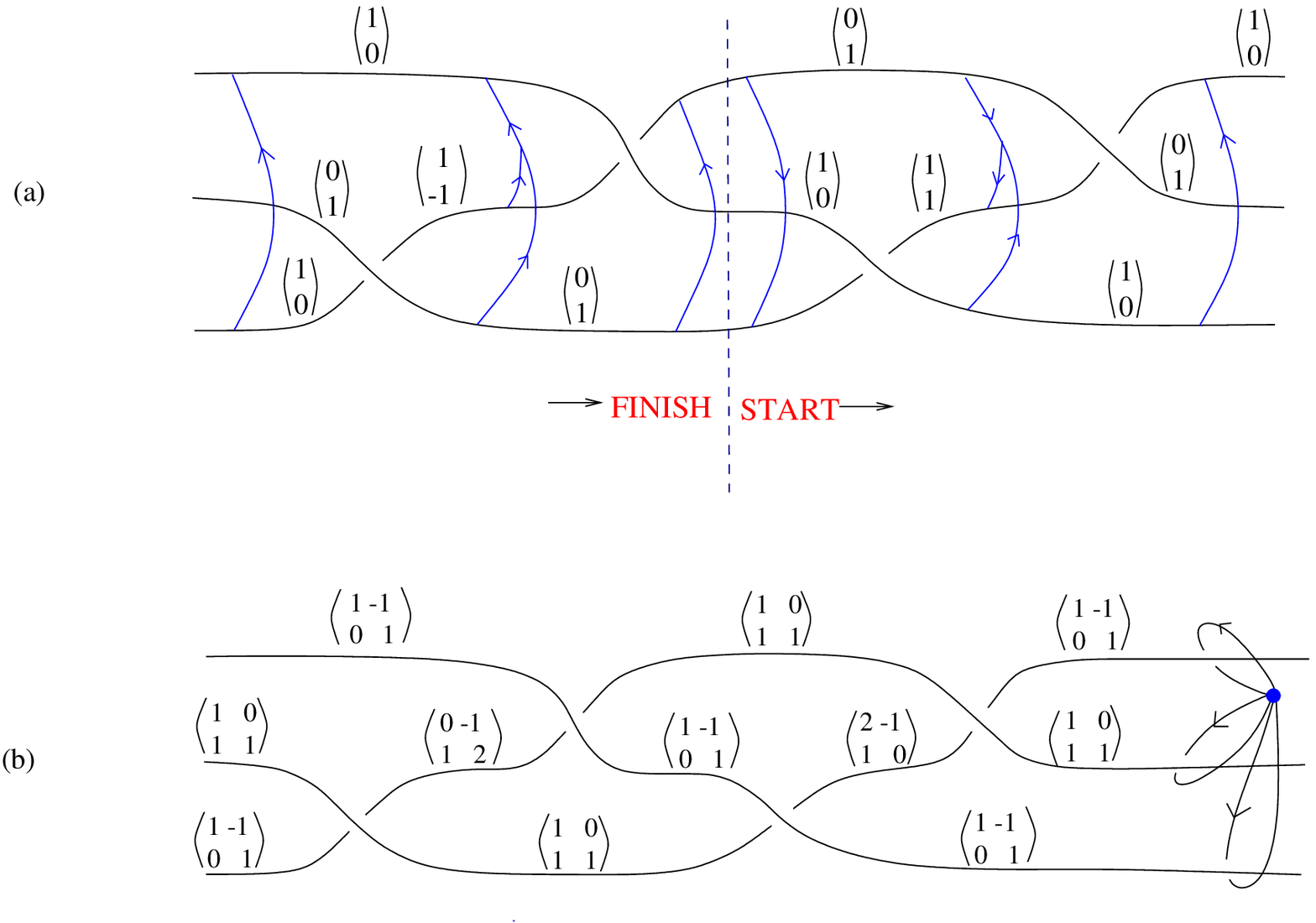}{(a) The root junction
$\vec{\alpha}$ can be transformed to $-\vec{\alpha}$ by a
translation all the way around the braid. As one moves from one
section of the braid to the next, Hanany-Witten transitions are
performed to keep the junction in a canonical form in the upper
half of a plane transverse to the braid.  The vanishing cycles
with respect to a path in the upper half plane are indicated above
each segment of the braid. The vanishing cycles are eigenvectors
of the monodromies indicated in (b).}

\onefigure{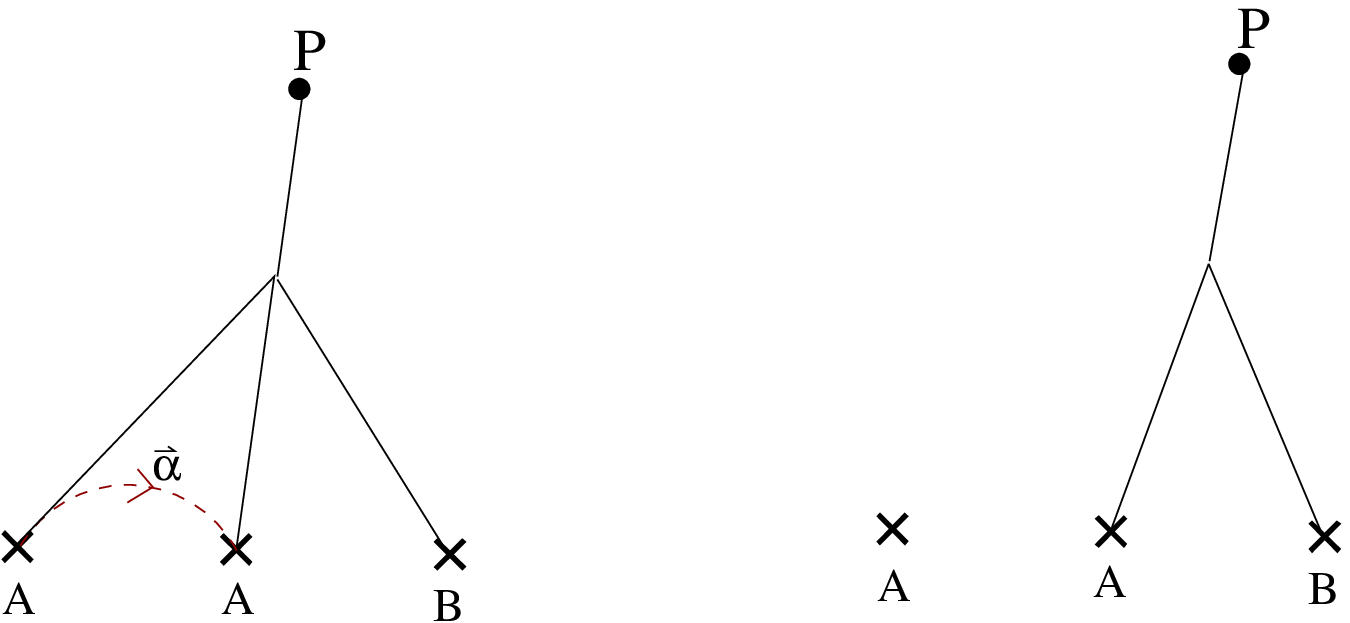}{In the three--strand braid representation
of the trefoil,  the root junction $\vec{\alpha}$ is equivalent to
zero. Therefore, any string junctions related by the addition or
subtraction of $\vec{\alpha}$ are equivalent. Thus, the string
junction lattice is actually that of a Kodaira type $II$,  rather
than of type $III$.}

The string junction lattice is only part of what is required to
determine the spectrum of light charged matter on a five--brane
wrapped near a singularity. Extra constraints are required, which
we will discuss later.   In this particular example, however,  we
expect that the infrared limit of the five-brane worldvolume
theory in the neighborhood of the cusp is the same as that in the
neighborhood of a type $II$ Kodaira fiber over a smooth component
of the discriminant curve.  This is because,  based on arguments
in \cite{aks},  the cusp singularity can not describe a conformal
fixed point with a Weierstrass form invariant under scale
transformations. Furthermore, the operators which deform a smooth
type $II$ into a cusp are irrelevant,  while those which deform a
smooth type $III$ into a cusp are relevant.  This is consistent
with the result that the  string junction lattice is that of a
type $II$.  We conclude that the light state spectrum at a cusp
singularity is $N=2$ supersymmetric and identical to the spectrum
for a Kodaira type $II$ fiber over points in the smooth part of
the discriminant curve.   This spectrum was computed in \cite{ggo,
bartontwo}.

\section{Deforming the equation: normal crossing intersections of the discriminant}\label{def}

  In the previous section, we gave an example of a singularity where it was not necessary
  to deform the discriminant curve, since the Kodaira type fibers in a neighborhood of
  $P_0$           are of type $I_1$.
Here we present  three different examples where the Kodaira type fibers in a neighborhood of
  $P_0$           are no longer of type $I_1$.  Therefore, in these examples,
  we first need to deform
the discriminant curve, before regularizing the singularity with
the sphere $S^3$.
  In the case of the normal crossing
intersection, the links  on $S^3$ are, in fact, simple linked
circles and there is an obvious minimal braid (with no extra
relations on the string junction lattice).
 Thus, while in the previous example we focused on the
 braids, in the following  examples we will focus on the deformations.

 Before we consider the examples of simple normal crossing singularities,
 we will revisit the case of a smooth part of the discriminant curve,
 as discussed in \cite{ggo}.
 Unlike \cite{ggo}, however, in this section we will use a deformation
 which leaves the multiplicity of the discriminant fixed at a point
 and regularize the singularity with the sphere $S^3$.
All the examples in \cite{ggo} could have been computed in this manner.
 We present this method here, since it gives a simple illustration of
 the features which will occur, in a more complicated form, for  simple
 normal crossing singularities.

\subsection{The smooth locus of the discriminant }\label{smooth}

  We consider the smooth part of the discriminant of Kodaira
  type $I_2$, in Figure 1.
Pick any point $P_0$ on the smooth part of $\sigma$, and choose the origin of the
$s,t$ coordinates to be at that point. In the neighborhood of this
point, the associated sections have the form
\begin{equation}
g_{2}=3a^2, \qquad g_{3}=a^3-s^2, \qquad \Delta= s^{2}f^{2}
\label{eq:81}
\end{equation}
where $a$ and $f$, to leading order in $s$, are non--zero
functions of $t$ only. (See for example, Table $1$ in Section \ref{notation}.

We choose a deformation of the Weierstrass representation which
fixes the order of vanishing (here the order of vanishing is $2$) of the discriminant at the origin.
After the deformation, the singular fibers outside
the origin become of Kodaira type $I_1$. A suitable such deformation of
the equation has the form
\begin{equation}
y^2=4x^3 -3a^2x -(a^3-(s+ \epsilon t)(s-\epsilon t))
\end{equation}
 with discriminant
\begin{equation}
\Delta_\epsilon= (s+ \epsilon t)(s-\epsilon t) b,
\end{equation}
where $b$ does not vanish near $P_0$.
 Note that this is also the ``relevant" deformation   used in \cite{ggo}
 of the general elliptic surface through $P_0$, with $\epsilon t$ the deformation
 parameter.

 The general elliptic surface of
the deformed model through $P_0$ (obtained by setting, say $t=0$)
 also has Kodaira type fiber of type $I_2$.
Similarly, the order of vanishing of the discriminant at $P_0$ is $2$,
  both in the deformed and non-deformed models.
 This satisfies the rules stated in Section \ref{methods}.
 Then, as for the cusp singularity, we will regulate the
crossing point by considering the intersection of the deformed discriminant
curve with the sphere $S^{3}$ of infinitesimal radius $R$ centered
$P_0$.
We obtain the  ``link'' shown in Figure \ref{s1} (see Appendix A
and A.3). Since this link is composed of points all in the smooth
part of the discriminant curve, the Kodaira type of fiber
degeneration over any point in the link can be determined by
canonical methods.

\onefigure{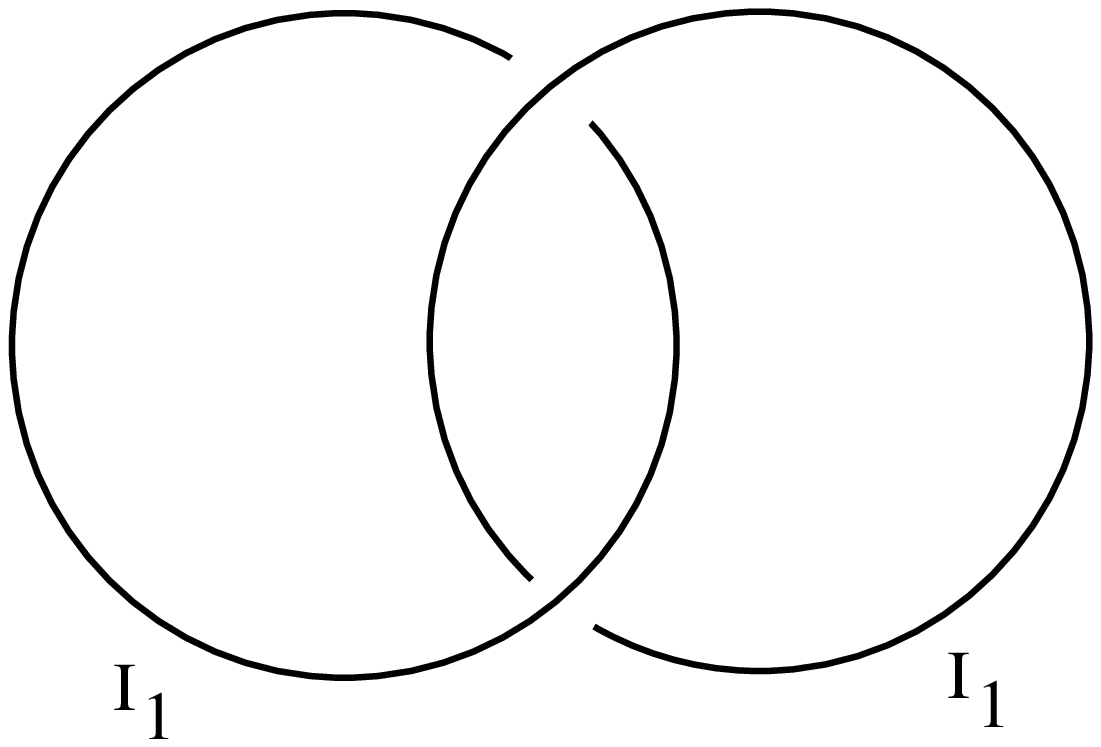}{Link associated to the deformed $I_2$ smooth
locus.}

As before, we represent this link as a braid.
Since this link is so simple, the  proof that the string junction lattice is that of the
minimal braid is straightforward.
 This {\it minimal braid} (see A.2) is shown in Figure $10$.

 \onefigure{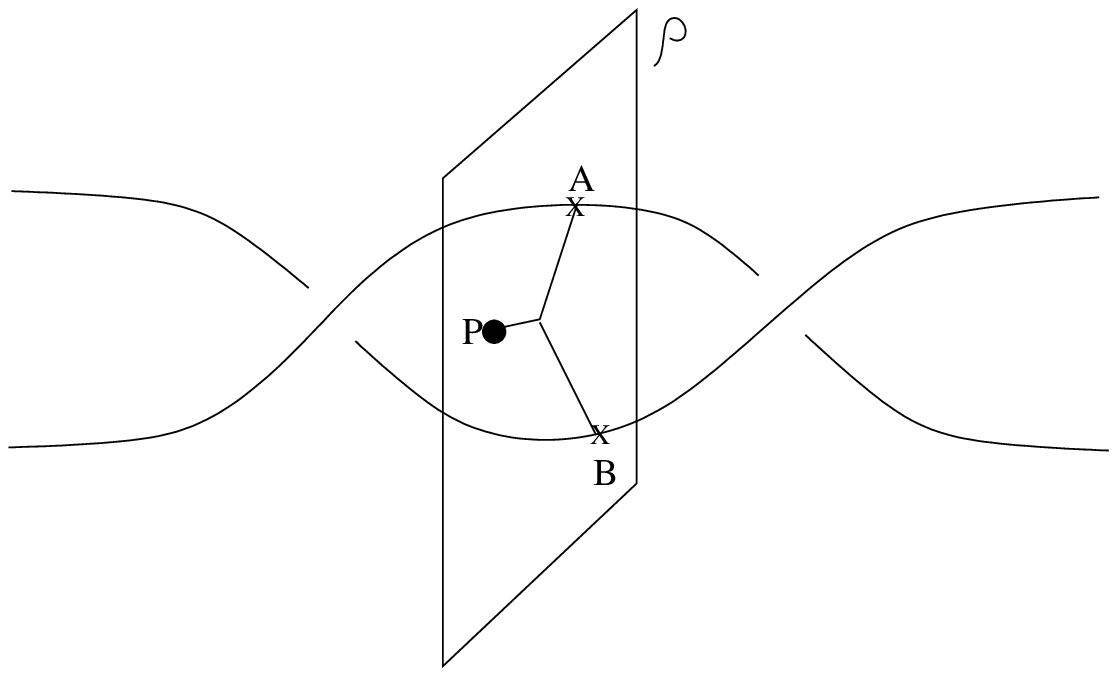}{Minimal braid associated to the deformed $I_2$ smooth locus. }

 Note
that the braid intersects $\cP$ in two points. The elliptic
fibration over $\cP$, which we denote by ${\cal{T}}_{R}$, is our
regularization of the surface ${\cal{T}}$. ${\cal{T}}_{R}$ has
two separated discriminant points,
 with Kodaira type $I_{1}$. The monodromies of these points can be computed and
are found to be of type $A$.
Using the Hanany--Witten mechanism, it is not too hard to show
that any string junction starting at $P$ can always be represented
by a two--legged string junction in the plane $\cP$. Note that the
string junction we obtain in this manner corresponds to a two
Kodaira type $I_{1}$ fiber degeneracy.

 Because the braid is composed by two simply linked segments,
 there are no further relations on the string junction lattice
 in the plane $\cal P$,  which is the lattice of a type $I_{2}$.
 Note that the vanishing of the discriminant at $P_0$
 is equal to
 the number of strands of the braid.
 This is a general fact,
 see A.2  and the  end
 of the Introduction.
Without the  restriction on the allowed deformations, we could
have  obtained a braid with fewer strands. It would no longer have
been  clear how to see a ``cut'' of this braid as a
deformation of the Kodaira type $I^2$.

 \medskip

 In general, if $P_0$ is a
smooth point of the discriminant, then any allowed deformation is
also a deformation of the general elliptic surface through $P_0$.
  The associated
 link is given by $n$ simple linked circles,
 where $n$ is the order of vanishing of the discriminant at $P_0$.
 The associated minimal braid has $n$ strands, because we have $n$
 distinct links; any
allowed deformation is also a deformation of the general elliptic
surface through $P_0$.  Hence, we find the string junction lattice of the
general surface through $P_0$.

\subsection{Simple Normal Crossing Intersection: Kodaira type $I_1$ and
$I_2$}\label{def1}

   We now apply the techniques described in the previous subsection
   to simple normal crossing intersection of the discriminant curve.
In the discriminant locus shown
in Figure $1$, there are $28$ simple normal crossing intersections
of the components $\Sigma$ and $\sigma$ of the curve.  Consider
one such intersection $P_0$ and choose local coordinates $s,t$
around $P_0$. Then, in the neighborhood of this point, the
sections $g_{2}$ and $g_{3}$ can be taken to be
\begin{equation}
g_{2}= 3a^2, \qquad  g_{3}= a^3 - s^2t, \label{eq:a}
\end{equation}
where $a$ is a suitable constant (see  Section \ref{notation}). The
equations of the Weierstrass model and the discriminant are given
by
\begin{equation}
y^{2}=4x^{3}- 3a^2x- (a^3 -s^2t) \label{eq:b}
\end{equation}
and
\begin{equation}
\Delta_\epsilon =s^2t [-2a^3 +s^2t]
\end{equation}
respectively.
 We choose a
deformation of the Weierstrass representation which fixes the
order of vanishing of the discriminant at the origin, while the
singular fibers outside the origin become Kodaira type $I_1$.

The equation of the deformed Weierstrass threefold and the
discriminant are given by
\begin{equation}
y^{2}=4x^{3}- 3a^2x- (a^3 -(s+\epsilon t)(s-\epsilon t))t
\label{eq:b}
\end{equation}
and
\begin{equation}
\Delta =(s+\epsilon t)(s-\epsilon t)t [-2a^3 +(s+\epsilon
t)(s-\epsilon t)t]
\end{equation}
respectively. We see that any point with Kodaira type $I_{2}$,
 is split into two
nearby discriminant points, each of Kodaira type $I_{1}$.
 Note
that the general elliptic surface of the deformed model through
$P_0$  (obtained, by setting, say, $t=s$) has the same type of
singularities as the non-deformed one and that the multiplicity of
the discriminant through $P_0$ is unchanged in the deformed and
non-deformed models.
 This satisfies the rules stated in Section \ref{methods}.
 Then, as in the previous example, we will regulate the
crossing point by considering the intersection of the deformed
discriminant curve with the sphere $S^{3}$ of infinitesimal radius
$R$ centered $P_0$.
We now  obtain the  ``link'' shown in Figure $11$. Again, this link
is composed of points all in the smooth part of the discriminant
curve and the Kodaira type of fiber degeneration over any point in
the knot can be determined by canonical methods.

\onefigure{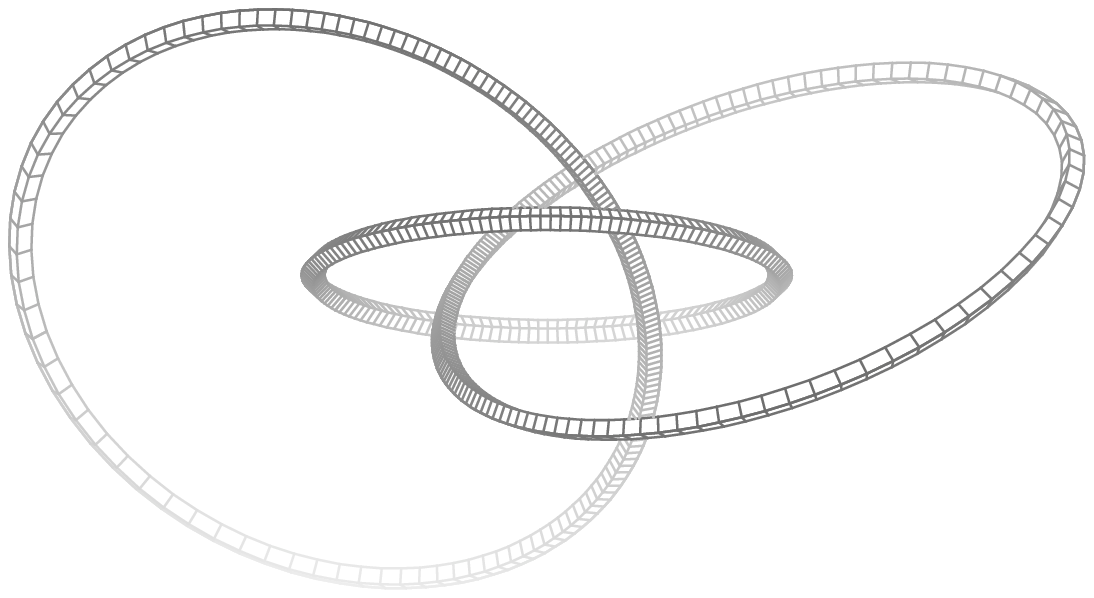}{Link associated to the deformed $I_2$-$I_1$
intersection.}

Next, we represent this link as a braid. Among the various braids
representations we will consider only the simplest braid
representation, which corresponds to choosing a generic base curve
for the intersecting surface.
 This {\it minimal braid} (see A.2) is shown in Figure $12$.

\onefigure{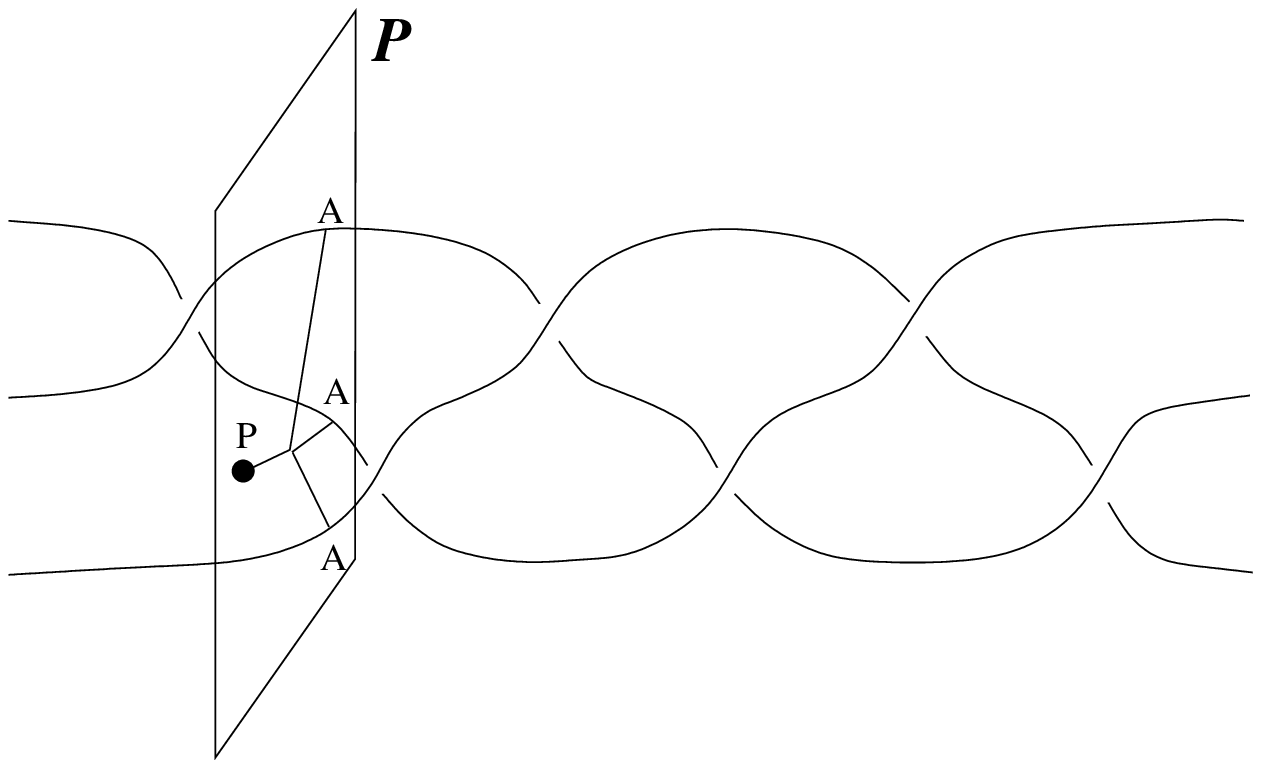}{Minimal braid associated to the deformed
$I_2$-$I_1$ intersection.  }

 Note
that the braid intersects $\cP$ in three points. The elliptic
fibration over $\cP$, which we denote by ${\cal{T}}_{R}$, is our
regularization of the surface ${\cal{T}}$. ${\cal{T}}_{R}$ has
three separated discriminant points,
 with Kodaira type $I_{1}$. The monodromies of these points can be computed and
are found to be of type $A$. This is indicated pictorially in Figure $12$.
Using the Hanany--Witten mechanism, it is not too hard to show
that any string junction starting at $P$ can always be represented
by a three--legged string junction in the plane $\cP$. As in the
previous example, because the braid does not induce any relations
among the end points, the associated string junction lattice  has monodromy structure
$A \cdot A \cdot A$. This is
the lattice of a type $I_{3}$ fiber.
 Therefore, the light state spectrum at the $I_1-I_2$ simple
 normal crossing identifies to that of a type $I_3$ fiber over a
 smooth part of the discriminant. This spectrum was presented in
 \cite{ggo}.

\subsection{Simple Normal Crossing Intersection: Kodaira type $I_1$ and
$I^*_0$}\label{def2}

Here we consider another type of simple normal crossing
intersection, which also occurs in case 1 of example 2 in
\cite{ggo}. Two different components $\Sigma$ and ${\cal{S}}$ of
the discriminant, of Kodaira type $I_1$ and $I^*_0$ respectively,
meet with a simple normal crossing intersection at a point $P_0$.
We can choose local coordinates $s,t$ around $P_0$, with $P_0$ as
the origin. Then, in the neighborhood of this point, the sections
$g_{2}$ and $g_{3}$ can be taken to be
\begin{equation}
g_{2}= 3(sa)^2, \qquad  g_{3}= (ac)^3 - s^3t, \label{eq:a}
\end{equation}
where $a$ is a suitable function (see Section \ref{notation}). The
equation of the deformed Weierstrass threefold and the
discriminant are given by
\begin{equation}
y^2= 4x^3 -3(sa)^2x+  (sa)^3 + t(s^3 )
\end{equation}
\begin{equation}
t \cdot (s^6) \cdot (4a^3 +t)=0.
\end{equation}
 As before, we choose a deformation of the Weierstrass representation
which fixes the singularity of the discriminant at the origin, so
that
 the singular fibers outside the origin become of type $I_1$ in the
 deformed model.
The equation of the deformed Weierstrass threefold and the
discriminant are given by
\begin{equation}
y^2= 4x^3 -3(sa)^2x+  (sa)^3 + t(s^3 - \epsilon ^3 t^3)
\end{equation}
\begin{equation}
t \cdot (s-\e t)\cdot ( s-\e e^{\frac{2\pi i}{3}}t)\cdot ( s- \e
e^{\frac{4\pi i}{3}}t) \cdot (-\e^3 t^4 + s^3(4a^3 +t))=0.
\end{equation}
This Weierstrass model defines a threefold $W_\epsilon$, which is
 smooth outside the point $y=x=t=s=0$ ($a$ is non-zero at the origin).
The singular fibers are of type $I_1$, except at the point
${s=t=0}$.
 This deformation preserves the type of singularity of the generic elliptic surface
through $P$. The corresponding braid can be shown to be a $7$-braid. It is more
complicated to draw, so we limit ourselves to describe it in
words, and display the (MAPLE generated) picture of the
corresponding knot. The braid is composed of four simple linked
circles (as in the previous example) and a 3-stranded minimal
braid.
 It is easy to see that,
by cutting with a general real plane $\cal P$,  we obtain a
monodromy structure of type $A^5 \cdot B \cdot C$. Since there are
no further relations among the different strands, this will also
be the monodromy structure in the threefold. From Table 4 in
\cite{ggo}, we find that the corresponding junction lattice is
that of a type $I^*_1$ singularity. Therefore, the light state
spectrum at the $I_1-I^*_0$ simple
 normal crossing identifies to that of a type $I_1^*$ fiber over a
 smooth part of the discriminant.

\onefigure{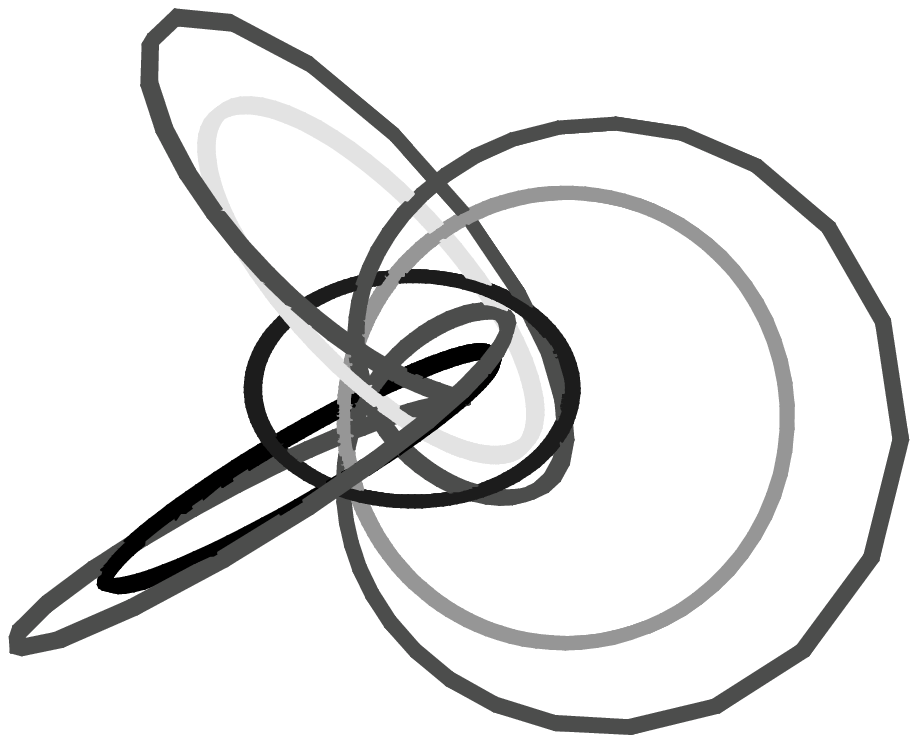}{Knot corresponding to the deformed $I^*_1$
and $I^*_0$ simple normal crossing intersection.}

\section{ A puzzle: Non-transversal Intersections}\label{susy1}

 We also studied many examples of non-transversal intersections.
We found that in these cases no deformation leaves the $I_1$ locus
fixed,
 and that we cannot determine
which strands of the knot correspond to the deformation of the
un-deformed points of the discriminant.
 Therefore, the two-dimensional analysis cannot be used
in these examples. This might indicate  that  supersymmetry is
broken  at the point $P_0$, and that the string junction lattice
of the threefold  no longer coincides with that
of the general elliptic surface through $P_0$, as in the  ${N} =2$ case.
As an example of this problem, let us consider a case where the
supersymmetry is known to be broken to  ${N}=1$.

In the discriminant locus shown in Figure $1$, there are $5$
non-transversal intersections of the components $\Sigma$ and $\cS$
of the curve. Let us consider one such intersection point and
choose local coordinates $s,t$ which vanish at that point. Then,
in the neighborhood of this point, the sections $g_{2}$ and
$g_{3}$ can be taken to be
\begin{equation}
g_{2}=st^{3}, \qquad g_{3}=t^{5} \label{eq:12}
\end{equation}
It follows that the Weierstrass representation of the fiber and
the discriminant are given by
\begin{equation}
y^{2}= 4x^{3}-st^{3}x-t^{5} \label{eq:13}
\end{equation}
and
\begin{equation}
\Delta=t^{9}(s^{3}-27t) \label{eq:14}
\end{equation}
respectively.
The Kodaira type over $t=0$ is of type $III^*$. A possible allowed
deformation is
\begin{equation}
y^{2}= 4x^{3}-st^{3}x-(t^{5} + \e s^5) \label{eq:13}
\end{equation}
with discriminant
\begin{equation}
\Delta=t^{9}s^{3}-27(t^{10}+\e^2 s^{10} +2\e t^{5}s^5 )
\label{eq:14}
\end{equation}
Because deformation does not leave the (original) $I_1$ locus fixed,
 we cannot determine
which strands of the knot correspond to the deformation of the
$III^*$ locus. We do not believe that any
such deformation exists. This makes examples of this kind completely
different from the $N=2$ cases discussed previously in this paper.

\section{Supersymmetry at Low Energy.}\label{susy2}

The field theory on the worldvolume of a five-brane in the bulk
has ${N}=1$ supersymmetry. However, the amount of supersymmetry
may be greater in the infrared limit. For example, if the
five-brane wraps an elliptic curve over a generic point in the
base of the Calabi-Yau threefold away from the discriminant curve,
the IR theory is an ${N} =4$ $U(1)$ gauge theory.  This can be
understood heuristically as a consequence of the correspondence
between long distances on the five-brane and short distances in
the transverse space-time. In the IR,  the five-brane probes only
the local transverse geometry of the Calabi-Yau base,  which is
$\RR^4$ when the brane wraps a fiber over a generic point in the
base. In the IR,  the scalar fields associated with this $\RR^4$,
together with the axion and $S_1/Z_2$ moduli, belong to a
multiplet under the ${N} =4$ $SO(6)$ R symmetry. The situation is
different when the five-brane wraps a fiber over the discriminant
curve. When this fiber lies over a smooth component of the
discriminant curve, the IR theory has ${N} =2$ supersymmetry.

When the fiber lies over a singular point of the discriminant
curve, infrared properties can be studied using the methods of
\cite{aks}.  The authors of \cite{aks}, studied the low energy
theories of three-brane probes of Calabi-Yau geometries in
F-theory,  which are the same as the theories which arise in our
context. If the IR theory is an interacting fixed point with ${N}
=1$ supersymmetry, then the local form of the singularity is
preserved by IR flow,  meaning that the local Weierstrass
equation,
\begin{equation}
y^2 = x^3 - f(s,t)x -g(s,t). \label{weier}
\end{equation}
is invariant under scaling when appropriate dimensions are
assigned to $x,y,u$ and $v$. The holomorphic three-form of the
Calabi-Yau threefold,
\begin{equation}
\Omega = \frac{ds \wedge dx \wedge dt}{2y}
\end{equation}
should have dimension $[\Omega]=2$ (see \cite{aks}). Furthermore,
for unitarity one requires $[s] > 1$ and $dim[t] > 1$.

As an example, consider the tangential crossing of $III^*$ with
$I_1$ described by
\begin{equation}
y^{2}= 4x^{3}-st^{3}x-t^{5},
\end{equation}
with a five-brane wrapped over the point $s=t=0$. This example
(discussed in \cite{aks}) is consistent with an ${N}=1$ CFT in the
infrared limit. The scaling relations $2[y] = 3[x] = [s] + 3[t] +
[x] = 5[t]$ and $[\Omega] = [s] + [t] + [x] - [y] = 2$ gives the
anomalous dimensions $[s] = 4/3$ and $[t] = 4$.

When the conditions for ${N}=1$ supersymmetry in the IR are not
satisfied, one expects the theory to flow to an ${N} =2$ fixed
point in the infrared limit.  Heuristically,  one can understand
this process as a smoothing of the singularity,  such that the
local geometry seen by the five-brane looks like that over a
smooth component of the discriminant curve.  As an example,
consider the cusp singularity $y^2 = 4x^3 -sx -t$, with a
five-brane wrapped over the point $s=t=0$. In the infrared limit,
the theory flows to a CFT with no dimensional parameters. If this
CFT has $N=1$ supersymmetry,  the discriminant curve singularity
should be fixed under the infrared flow,  that is the Weierstrass
equation above should have no dimensional parameters. The
dimensions would have to satisfy the following relations:  $2[y] =
3[x] = [s] + [x] = [t]$, and also that  $[\Omega] = [s] + [t] +
[x] - [y] = 2$ . This would imply that $[s] = 8/9$, which is not
consistent with the unitarity bound $[s] \ge 1$ on scalar fields
in a unitary superconformal theory. Thus the cusp can not describe
a non-trivial $N=1$ fixed point. A natural guess is that the
theory flows in the infrared to an $N=2$ conformal theory,
described either by the Weierstrass model $y^2 = 4x^3 - t$ or by
$y^2 = 4x^3 - sx$. The former possibility seems more likely for
the following reasons.  In the $N=2$ CFT described by $y^2 = 4x^3
-t$, the scaling dimension of $[s]$ is $1$,  as $s$ is a free
scalar field describing motion parallel to the discriminant curve.
The scaling relation $2[y] = 3[x] = [t]$,  together with $[\Omega]
= [s] + [t] + [x] - [y] = 2$ yields $[t] = 6/5$.  Then the
(analytic) deformation which gives a cusp singularity is given by
$y^2 = 4x^3 - \epsilon sx -t$.  This deformation is irrelevant
since $[\epsilon] = -1/5$. However, for the $N=2$ CFT described by
$y^2 = 4x^3 - sx$, the scaling dimensions satisfy $[t]=1$,  $2[y]
= 3[x] = [s] + [x]$ and $[\Omega] = [s] + [t] + [x] -[y] = 2$.
This means that the deformation which gives a cusp singularity,
$y^2 = 4x^3 - sx - \epsilon t$, is relevant since $[\epsilon] =
1$. This suggests that in the infrared limit,  the theory
described by the cusp singularity flows to an $N=2$ CFT described
by the Weierstrass model of a type $II$ Kodaira fiber over a
smooth component of the discriminant curve.

As one further example, consider the simple normal crossing of
$I_2$ with $I_1$ described by $y^{2}=4x^{3}- 3a^2x- (a^3-s^2t)$,
with a five-brane wrapped over $s=t=0$. There is no way to assign
scaling dimensions, so that there are no dimensional parameters in
the Weierstrass model. Hence $[s]$ and $[t]$ have dimension $>1$,
and $[\Omega] =2$. One would have to satisfy $2[y] = 3[x] = 2[a] +
[x] = 3[a] = 2[s] + [t]$, but if the parameter $a$ has dimension
zero, then $2[s] + [t] = 0$ and $[\Omega] = [s] + [t] =2$. Note
that, in this example,  the $I_2$ and $I_1$ fibers are mutually
local (same vanishing cycle),  so that there are no states with
mutually non-local $(p,q)$ charges.  Consequently one does not
expect to find a non-trivial CFT in the infrared limit.  The IR
flow will never give a local form of the Weierstrass equation
which is invariant under scaling. The low energy theory is a
$U(1)$ gauge theory with electrically charged matter,  and flows
to a free theory in the infrared limit.

\vskip 0.2in {\bf Acknowledgements} We would like to thank B.
Johnson, P. Melvin and L. Zulli for helpful conversations.

 B. Ovrut is
supported in part by  the DOE under contract No.
DE-ACO2-76-ER-03071. A. Grassi's research was supported in part by
NSF grants DMS-9706707 and DMS-0074980. Part of this work was
completed while A.G. was a member at the Institute for Advanced
Study, Princeton, NJ and supported by N.S.F. grant DMS-9729992. Z.
Guralnik is supported in part by  the DOE under contract No.
DE-ACO2-76-ER-03071.

\section*{Appendix: Curve Singularities, Knots and Braids}

In this section, we derive examples of knots (or links) and braids
associated to an isolated plane curve singularity. In
\cite{milnor} Milnor studied properties of isolated (complex)
singularities of hypersurfaces in $\CC ^n=\RR ^{2n}$ by
intersecting the hypersurface with a  sphere $S^{2n-1}$  of radius
$R$ centered at the singularity. In the case of complex  plane
curves, the intersection is a real curve, called a knot, if it has
 one component, a link otherwise.
 These are also called {\it algebraic knots} (links).
The idea is that the type of singularity of the curve is closely
related to the topological property of the associated knot (or
link). For example, as we will see below, the knot associated with
a smooth point is a circle (unknot).

Take $(s,t)$ to be complex coordinates in the plane $\CC ^2 \sim
\RR^4$, so that the complex curve has an isolated singularity at
the point ${(s,t)=(0,0)}$. To describe the intersection of the
curve and $S^3$ it is easier to use polar coordinates:
$$
\xi e^{i \theta}=t, \eta e^{i \phi}=s;
$$%
the equation of the sphere of radius $R$ is then $\xi ^2 + \eta
^2=R^2$. Recall that $S^3$ can also be thought of two solid tori
glued along the boundary. In the example we consider, the
intersection of the complex curve with $S^3$ will be a knot (or
link) on the (boundary of) tori. These are called {\it torus}
knots (or links).

\subsection*{A.1: A smooth point is the unknot}\label{app1}

Let us assume that ${s=0 }$ is the local equation of the curve
around the origin. It is easy to see that the intersection of the
curve with the sphere $\xi ^2 + \eta ^2=R^2$ is a circle $S^1$ of
radius $R$.
 We can cut the circle in a point and obtain a line segment; we can recover the circle by identifying
the endpoints of the segment. The segment is called an associated
{\it braid}. Similarly, given a braid, we obtain a knot by  {\it
closing} the braid.
 For a precise definition of a braid see, for example, \cite{burde} and \cite{murasugi}.

\subsection*{A.2: On braids}\label{appb}

 We can associate many different braids to a knot; it is a hard result that
every knot (link) can be represented by a closed braid
\cite{alexander}; (see again  \cite{burde} and \cite{murasugi}).
 Many different braids can be associated to a knot (link) \cite{burde} (2.14).
Among the different braid representations of the knot, the one
with the fewest strands is called the {\it minimal braid}. The
number of strands of the minimal braid is called the {\it braid
index}.
 The braid index is an invariant of the knot \cite{burde}.
Our analysis of string junction lattices led us to conjecture that
the braid index of these algebraic knot would always be equal to
the order of vanishing of the complex curve at $(s,t)=(0,0)$.
 It turns out that this statement is indeed
always true, but it is a hard and beautiful result. The first
proof was given in \cite{schubert}. A shorter, elegant proof is
given by \cite{williams}, using technique from dynamical system
(see also \cite{libgober} in the same volume, for an algebraic
version). In the above example, for the smooth point (i.e. the
order of vanishing is one),
 we have a one strand braid.

\subsection*{A.3: An isolated cusp singularity and the trefoil knot}\label{app2}

As in Section \ref{cusps}, we take $ t^2 + s^3=0 $ to be the local
equation of the singularity. (Note that the order of vanishing is
two, and we will find a two-strand minimal braid). We show that we
obtain the trefoil knot, and we construct two braids, the minimal
two-strand braid and a three-stranded braid. It is easy to see
that the intersection of the discriminant and the sphere $S^3_R$
of radius $R$ centered at the origin is the parametric
 curve:
 $$(t,s)= (\xi_0 e ^{i3\theta}, \eta_0 e^{i \frac{\pi}{3}} e^{i2\theta}),$$
where  $\xi _0$ and $\eta_0$ are fixed positive constant which
depend on $R$. From the parametric equation we see that this real
curve lies on $S^1 \times S^1$, the product of two circles of
radius $\xi_0$ and $\eta_0$ respectively; the exponential $( e
^{i3\theta}, e^{i2\theta})$ give a slope of $(3,2)$ on the square
obtained by ``cutting open'' the torus (see Figure
\ref{cuspsontours}).

\onefigure{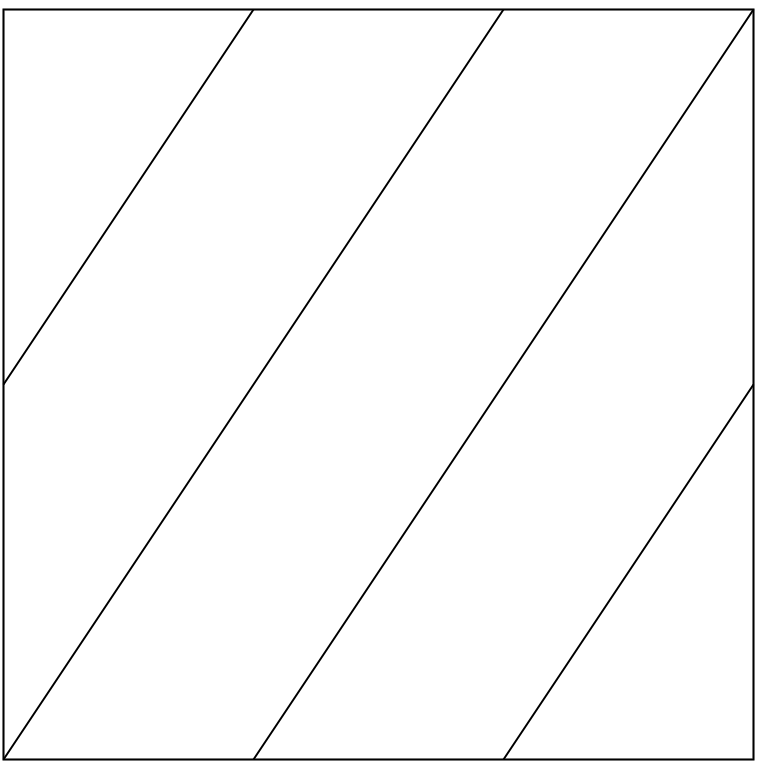}{The torus is obtained by identifying the
top  with the bottom and the left with the right  side. The
horizontal line is the $t$ axis ($s=0$), while the vertical one is
the $s$ axis ($t=0$). The slanted line represents the intersection
of the complex curve of equation $ t^2 + s^3=0 $ and $S^3$ of
slope $(3,2)$. Note that a vertical cut corresponds on the plane
$(s,t)$ with the line $s$ constant, and an horizontal cut with the
line $t$ constant.  } \vskip 0.3in

 This curve is connected, hence it is a {\it knot}; it is often denoted
  as the  $(3,2)$ {\it  torus knot}.
In Figure \ref{treontorus} we can see that this knot is exactly
the trefoil knot of Figure \ref{trefoil} in Section \ref{cusps}.

A general result shows that  all the knots obtained by
intersecting a complex curve with $S^3$
 are torus knots or can be described from torus knots (the
 ``iterated torus knots'').

\onefigure{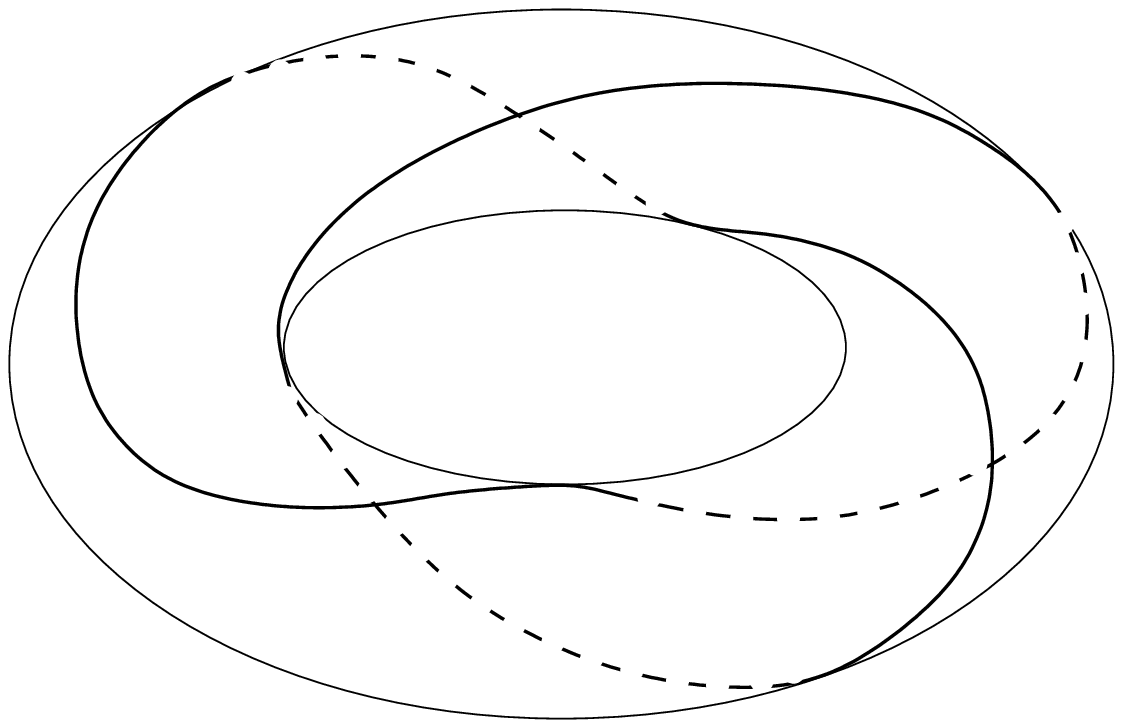}{The $(3,2)$ knot is the trefoil.}

In the next Figure \ref{cuspinbraids} we show that by ``folding''
the square in Figure
 \ref{cuspsontours}
horizontally (vertically), we obtain a two (three)-strand braid.
These foldings correspond to cutting the torus (and the knot)
along a horizontal (resp vertical) meridian.

\onefigure{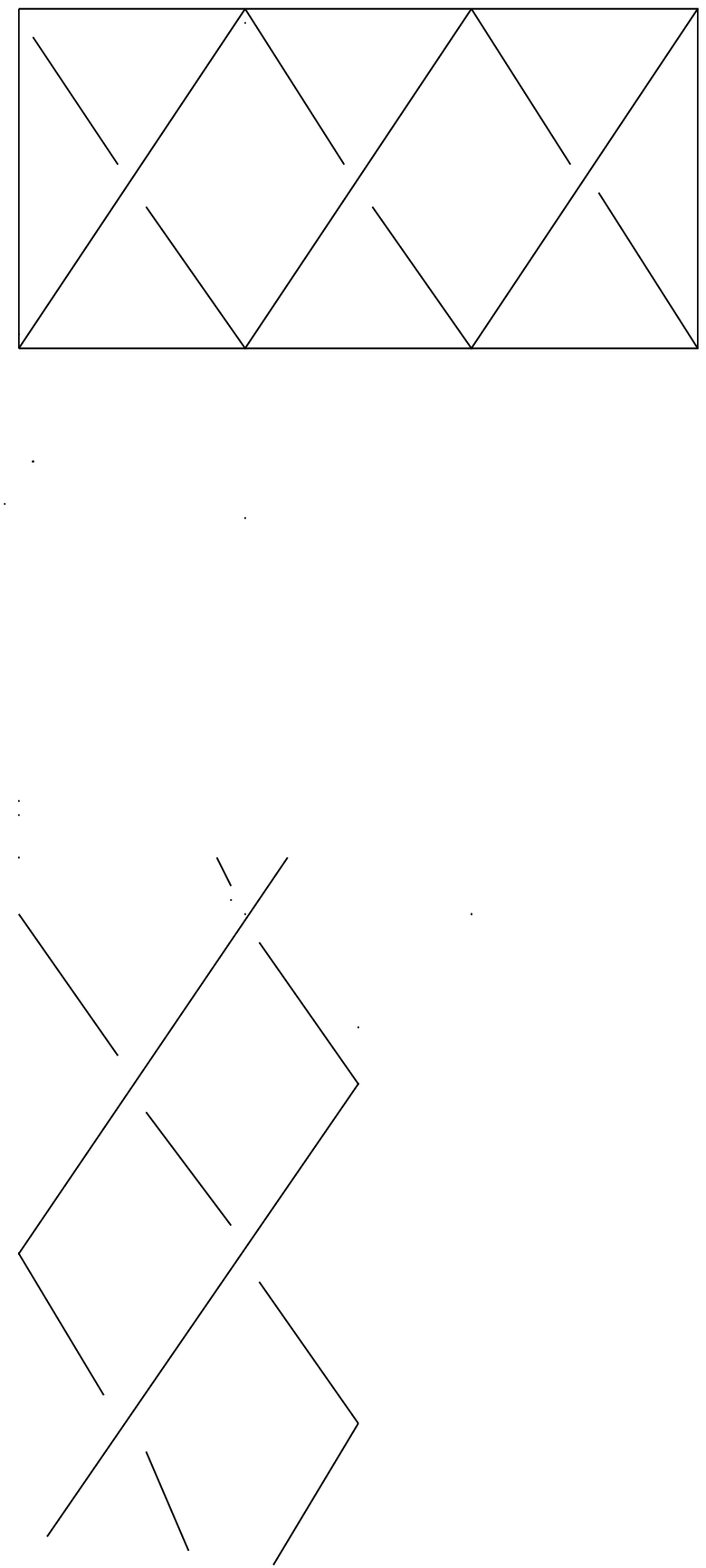}{ Two braids representations of the
$(3,2)$ torus knot. The two braids are obtained by cutting the
torus along the meridian and the longitude. This corresponds by
folding the square in the previous pictures in half, first
horizontally then vertically.}
\newpage
It is easy to see that the first braid corresponds to a cut
$t=c_1$ and the second to a cut $s=c_2$, where $c_1$ and $c_2$ are
constants depending on $R$, the radius of $S^3$. In the limit $R
\to 0$, the first $s=c_1$ cut intersects the cuspidal curve with
multiplicity $2$ (a generic intersection), while  the second cut
$s=c_2$  has intersection multiplicity $3$.

\subsection*{A.4: A transversal intersection (nodal singularity)}\label{app3}

Here we consider a local equation $(s+ \e t)(s- \e t)=0$, as in
Section \ref{smooth}; instead of a knot, we will obtain a link. It
is easy to see the intersection of the curve with $S^3$ lives
again on $S^1 \times S^1$. In the following Figure  \ref{s2} we
represent this torus by a square with identified sides, and the
braid obtained by folding horizontally (as in the previous
section).
 We see that we have two differen simply linked components.

\onefigure{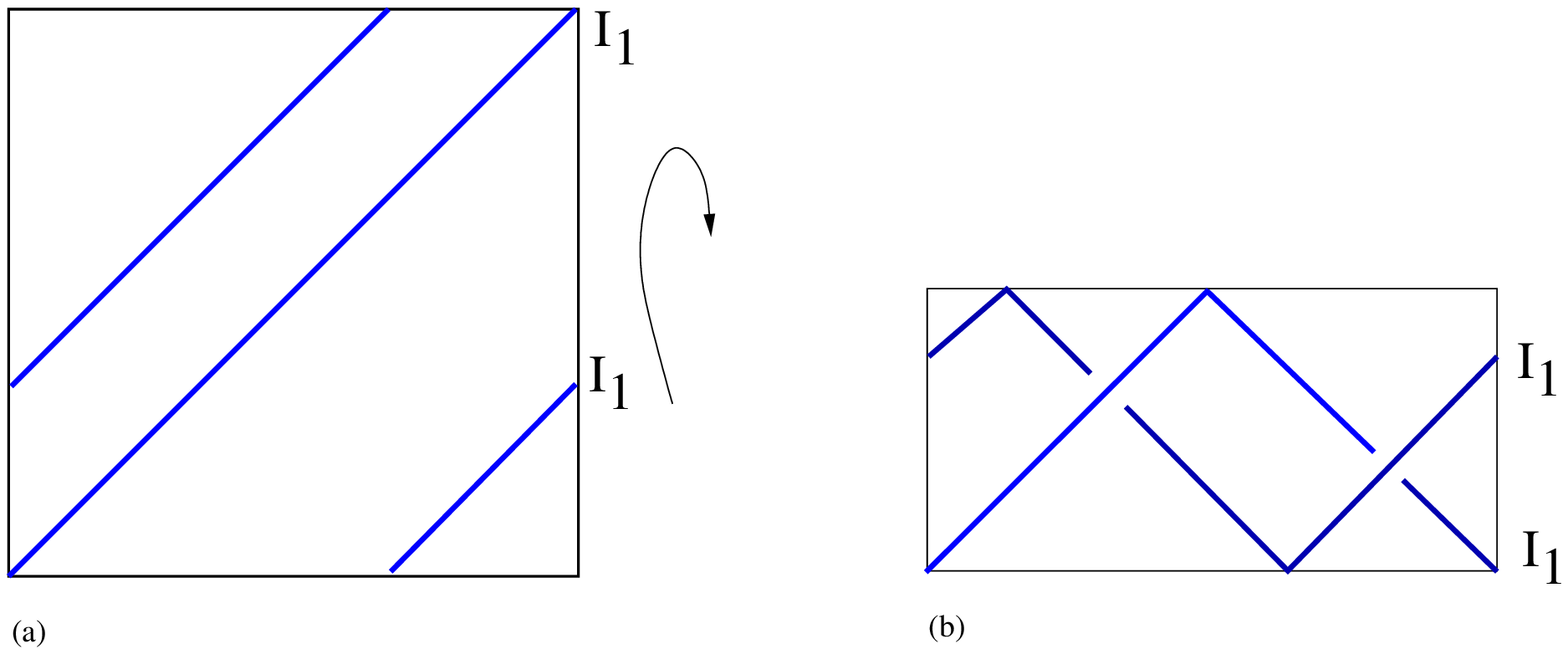}{The link associated with a nodal singularity and a
two-strand braid.}


\begin{thebibliography}{99}



\bibitem{losw1} A. Lukas, B.~A. Ovrut, K.~S. Stelle and D. Waldram,
    {\em The Universe as a Domain Wall},
    Phys.Rev. D59 (1999) 086001; {\em Heterotic M-theory in Five Dimensions},
    Nucl.Phys. B552 (1999) 246-290.



\bibitem{nse} A. Lukas, B.~A. Ovrut and D. Waldram,
    {\em Non-Standard Embedding and Five-Branes in Heterotic M-Theory},
    Phys.Rev. D59 (1999) 106005.



\bibitem{don1} R. Donagi, A. Lukas, B.~A. Ovrut and D. Waldram, {\em
    Non-Perturbative Vacua and Particle Physics in M-Theory},
    {\em JHEP} 9905 (1999) 018; {\em
    Holomorphic Vector Bundles and Non-Perturbative Vacua in M-Theory},
    {\em JHEP} 9906 (1999) 034.




\bibitem{fbs} A. Lukas, B. A. Ovrut and  D. Waldram, {\em
    Five--Branes and Supersymmetry Breaking in M--Theory},
    {\em JHEP} 9904 (1999) 009.



\bibitem{ms} R. Donagi, B. A. Ovrut and  D. Waldram, {\em
    Moduli Spaces of Fivebranes on Elliptic Calabi-Yau Threefolds},
    {\em JHEP} 9911 (1999) 030.



\bibitem{ppm} R. Donagi, B. A. Ovrut, T. Pantev and  D. Waldram, {\em
    Standard Models from Heterotic M-theory},
    hep-th/9912208.



\bibitem{si} B. A. Ovrut, T. Pantev and  J. Park, {\em
    Small Instanton Transitions in Heterotic M-Theory},
    hep-th/0001133.



\bibitem{HW1} P. Ho\v rava and E. Witten, {\em
    Heterotic and Type I String Dynamics from Eleven Dimensions},
    {\em Nucl. Phys.} {\bf B460}(1996) 506; {\em Eleven--Dimensional
    Supergravity on a Manifold with Boundary},
    {\em Nucl. Phys.} {\bf B475}(1996) 94.



\bibitem{ggo} A. Grassi, Z. Guralnik and B. Ovrut, {\em Five-Brane BPS States
    in Heterotic M-theory}, hep-th/0005121.



\bibitem{aks} O. Aharony, S. Kachru and E. Silverstein, {\em New $N=1$
    Superconformal Field Theories in Four Dimensions from D-brane Probes},
    Nucl.Phys.B488 (1997) 159,  hep-th/9610205.



\bibitem{dim} N. Arkani-Hamed, S. Dimopoulos and G. Dvali,
    {\em Phys. Lett.} {\bf B429} (1998) 263;
    I. Antoniadis, N. Arkani-Hamed, S. Dimopoulos and G. Dvali,
    {\em Phys. Lett.} {\bf B436} (1998) 257;
    L. Randall and R. Sundrum, {\em An Alternative to
    Compactification}, Phys.Rev.Lett. 83 (1999) 4690-4693.



\bibitem{kod} K. Kodaira, {\em On compact analytic surfaces, II, III},
Ann. of Math. {\bf 77} (1963) 563--626, {\bf 78} (1963) 1--40.





\bibitem{cosm} A. Lukas, B. A. Ovrut, K.S. Stelle and D. Waldram {\em
    Cosmological Solutions of Horava-Witten Theory},
    Phys.Rev. D59 (1999) 086001.



\bibitem{cosm1} A. Lukas, B. A. Ovrut, and D. Waldram {\em
    Boundary Inflation},
    Phys.Rev. D61 (2000) 023506.



\bibitem{cosm2} M. Braendle, A. Lukas and B. A. Ovrut {\em
    Heterotic M-Theory Cosmology in Four and Five Dimensions},
    hep-th/0003256.



\bibitem{cosm3} G. Huey, P.J. Steinhardt, B. A. Ovrut, and D. Waldram {\em
    A Cosmological Mechanism for Stabilizing Moduli},
    Phys.Lett. B476 (2000) 379-386.



\bibitem{cosm4} A. Lukas, B. A. Ovrut, and D. Waldram {\em
    Cosmological Solutions of Type II String Theory},
    Phys.Lett. B393 (1997) 65-71; {\em String and M-Theory Cosmological
    Solutions with Ramond Forms}, Nucl.Phys. B495 (1997) 365-399;
    {\em Stabilizing dilaton and moduli vacua in string and M--Theory
    cosmology}, Nucl.Phys. B509 (1998) 169-193.



\bibitem{FMW} R. Friedman, J. W. Morgan, E. Witten, {\em
    Vector Bundles and F Theory},
    Commun.Math.Phys. 187 (1997) 679-743.



\bibitem{cur} G. Curio, {\em Chiral matter and transitions in heterotic
    string models}, {\em Phys. Lett.} {\bf B435} 39 (1998).



\bibitem{ba} B. Andreas, {\em On Vector Bundles and Chiral Matter in $N=1$
   Heterotic Compactifications}, hep-th/9802202.



\bibitem{na} N. Nakayama, {\em On Weierstrass models}, in: ``Algebraic
Geometry and Commutative Algebra,'' Vol. II, Kinokuniya, Tokyo,
1988, pp. 405--431.



\bibitem{g} A. Grassi, {\em On minimal models of elliptic threefolds},
{\em Math. Ann.} {\bf 290} (1990) 287-301.



\bibitem{ar} V. Braun, P. Candelas, X. De la Ossa, A. Grassi
{\em Dualities  between $N=1$ theories}, hep-th/0001208.



\bibitem{GM} A.~Grassi and D.~R.~Morrison {\em Group representations and the
Euler characteristic of elliptically fibered Calabi--Yau
threefolds}.




\bibitem{BpvV} W. Barth, C. Paters, A. Van de Ven ,
``Compact Complex Surfaces,''  Ergebn. Math.  Grenzegeb. (3) {\bf
4}, Springer-Verlag, Berlin, 1984.




\bibitem{Yau} B. Greene, A, Shapere, C. Vafa, S. Yau {\it Stringy
cosmic strings and non-compact Calabi-Yau manifolds} Nucl.~Phys.
{\bf B 337}
  (1990) 1.




\bibitem{bartonone} O.~DeWolfe and B.~Zwiebach, {\it String junctions for
    arbitrary Lie algebra representations} Nucl.~Phys. {\bf B541}
  (1999) 509-565, hep-th/9804210.


\bibitem{bartontwo} O.~DeWolfe, T.~Hauer, A.~Iqbal and B.~Zwiebach, {\it
    Constraints On The BPS Spectrum Of N=2, D=4 Theories With A-D-E
    Flavor Symmetry}, Nucl.~Phys. {\bf B534} (1998) 261-274,
  hep-th/9805220.


\bibitem{sethi} A.~Mikhailov, N.~Nekrasov, S.~Sethi, {\it Geometric
    Realizations of BPS States in N=2 Theories}, Nucl.~Phys. {\bf
    B531} (1998) 345-362, hep-th/9803142.



\bibitem{ferrari1}F. ~Ferrari, {\it The Dyon Spectra of Finite Gauge
Theories,}
                Nucl.Phys.B501 (1997) 53-96, hep-th/9702166.



\bibitem{ferrari2}F. ~Ferrari, A. ~Bilal, {\it The Strong Coupling
        Spectrum of the Seiberg Witten Theory,}
        Nucl.Phys.B469 (1996) 387-402, hep-th/9602082.




\bibitem{argyresdouglas} P.~Argyres and M.~Douglas,
        {\it New Phenomena in $SU(3)$ Supersymmetric Gauge Theory},
        Nucl.~Phys. {\bf B448} (1995) 93-126, hep-th/9505062.



\bibitem{argwitten} P.~Argyres, M.~Plesser, N.~Seiberg and E.~Witten,
        {\it New $N=2$ Superconformal Field Theories in Four Dimensions},
        Nucl.~Phys. {\bf B461} (1996) 71-84, hep-th/9511154.



\bibitem{minahan} J.~Minahan and D.~Nemeschansky,
        {\it An $N=2$ Superconformal Fixed Point with $E(6)$ Global
        Symmetry},
        Nucl.~Phys. {\bf B482} (1996) 142-152, hep-th/9608047;\hfil\break
        J.~Minahan and D.~Nemeschansky,
        {\it Superconformal Fixed Points with $E(N)$ Global Symmetry.}
        Nucl.~Phys. {\bf B489} (1997) 24-46, hep-th/9610076.



\bibitem{bershadskyetal} M. Bershadsky, K. Intrilligator, S. Kachru, D. R.
Morrison,
        V. Sadov, and C. Vafa, {\it Geometric Singularities and Enhanced Gauge
        Symmetries}, Nucl.Phys. {\bf B481} (1996) 215-252, hep-th/9605200.



\bibitem{aspinwallkatzmorrison} P. Aspinwall, S. Katz, and D. R. Morrison,
{\it
        Lie Groups, Calabi-Yau Threefolds, and F-Theory}, hep-th/0002012.




\bibitem{milnor} J. Milnor
``Singular points of compex hypersurfaces" Ann. Math. Studies {\bf
61}, Princeton, NJ: Princeton University Press, 1968



\bibitem{burde} G. Burde, H. Zieschang ``Knots", de Gruyter Studies in Math, vol. 5,
Berlin-New York 1985.



\bibitem{alexander} J.W. Alexander {\it A lemma on systems of knotted curves} Proc. Nat. Acad.
Sci. USA, {\bf 9} 1923, 93--95




\bibitem{schubert} H. Schubert {\it Uber eine numerische Knoteninvariante} Math. Z. 61, 245--288,
1954



\bibitem{williams} R. Williams {\it The braid index of an algebraic link} in ``Braids", Santa
Cruz, 1986, 607--703



\bibitem{libgober} A. Libgober {\it On divisibility properties of braids associated with
algebraic curves} in ``Braids", Santa Cruz, 1986, 607--703



\bibitem{kawauchi} A. Kawauchi
``A survey of knot theory", Birkhauser, Basel, 1996



\bibitem{murasugi} K. Murasugi ``Knot theory and its applications"
 Birkhäuser, Boston, 1996




\end{thebibliography}
\end{document}